\documentclass[12pt]{article}

\usepackage{color}
\usepackage{amsmath,amssymb}
\usepackage{latexsym}
\usepackage[dvips]{graphicx,psfrag}
\usepackage{cite}
\usepackage{wrapfig}

\allowdisplaybreaks[1]

\newcommand{\eq}[1]{(\ref{#1})}
\newcommand{\nn}{\nonumber}
\newcommand{\ds}{\displaystyle}

\newcommand{\vev}[1]{\left\langle #1 \right\rangle}
\newcommand{\ket}[1]{\bigl|#1\bigr>}
\newcommand{\bra}[1]{\bigl<#1\bigr|}
\newcommand{\bracket}[2]{\left.\left\langle #1\right|#2\right\rangle}
\newcommand{\del}{\partial}

\newcommand{\bP}{\boldsymbol{P}}
\newcommand{\bQ}{\boldsymbol{Q}}

\newcommand{\bXi}{\boldsymbol{\Xi}}
\newcommand{\bPi}{\boldsymbol{\Pi}}

\DeclareMathOperator{\tr}{tr}

\renewcommand{\thefootnote}{\fnsymbol{footnote}}

\makeatletter
\@addtoreset{equation}{section}

\begin{document}


\begin{titlepage}
\thispagestyle{empty} 
\begin{flushright}
arXiv:1003.1626 \\
\vspace{0.2cm}
March 2010 
\end{flushright}

\vspace{2.3cm}

\begin{center}
\noindent{\large \textbf{
Fractional-Superstring Amplitudes,\\ 
\vspace{.43cm}
Multi-Cut Matrix Models 
and Non-Critical M Theory
}}
\end{center}

\vspace{1cm}

\begin{center}
\noindent{Chuan-Tsung Chan\footnote{ctchan@thu.edu.tw}$^{,p}$, 
Hirotaka Irie\footnote{irie@phys.ntu.edu.tw}$^{,q}$ 
and Chi-Hsien Yeh\footnote{d95222008@ntu.edu.tw}$^{,q}$}
\end{center}
\vspace{0.5cm}
\begin{center}
{\it 
$^{p}$Department of Physics, Tunghai University, Taiwan, 40704\\
\vspace{.3cm}
$^{q}$Department of Physics and Center for Theoretical Sciences, \\
National Taiwan University, Taipei 10617, Taiwan, R.O.C 
}
\end{center}

\vspace{1.5cm}

\begin{abstract}
Multi-cut two-matrix models are studied in the $\mathbb Z_k$ symmetry breaking 
$k$-cut $(\hat p,\hat q)$ critical points which should correspond to $(\hat p,\hat q)$ 
minimal $k$-fractional superstring theory. 
FZZT-brane or macroscopic loop amplitudes are obtained in all of these critical points 
and found to have two kinds of solutions in general.  Each of these solutions is expressed by 
hyperbolic cosine or sine functions with proper phase shifts. 
The algebraic geometries and ZZ-brane disk amplitudes (instanton actions) 
of these solutions are also studied. 
In particular, our results suggest that minimal $\infty$-fractional superstring theory 
can be viewed as a mother theory which includes all the minimal $k$-fractional superstring theories 
$(k=1,2,\cdots)$ as its perturbative vacua in the weak-coupling string landscape. 
Our results also indicate that, in the strong coupling regime of this fractional superstring 
theory, there is a three-dimensional theory which would be understood 
as the non-critical version of M theory in the sense proposed by 
P.~Ho\v rava and C.~A.~Keeler. 
\end{abstract}

\end{titlepage}

\newpage

\renewcommand{\thefootnote}{\arabic{footnote}}
\setcounter{footnote}{0}


\section{Introduction and summary}

Non-critical string theory \cite{Polyakov} provides a simple and tractable toy model 
to understand various aspects of string theory. This string theory is solvable not only 
in the world sheet CFT approach \cite{KPZ,DDK,DOZZ,Teschner,FZZT,ZZ,sDOZZ,fuku-hoso}, 
but also in the non-perturbative 
formulation based on matrix models \cite{DSL,TwoMatString,GrossMigdal2,fkn,fy12,fy3}.  
The solvability of the theory has uncovered many non-perturbative behaviors of string theory 
beyond string-coupling perturbative expansions \cite{Shenker} 
which played an important role in the discovery of D-branes \cite{Polchinski}. 
This theory has also provided a simple toy model of gauge/string correspondence \cite{Reloaded,Paradigm} 
and led to a new correspondence between two-cut critical points 
\cite{GrossWitten,PeShe,DSS,Nappi,CDM,HMPN} 
of the matrix models and type 0 superstrings \cite{TT,NewHat,UniCom}. 
Along this line, multi-cut extension of the correspondence has also been proposed in \cite{irie2}: 
the infinite sequence of correspondences between $k$-cut matrix models \cite{MultiCut} 
and $k$-fractional superstring theory \cite{FSST}.%
\footnote{In this paper, the $k$-th fractional superstring theory is called $k$-fractional superstring theory. 
Therefore, on the worldsheets of the $k$-fractional superstrings, 
there are $\mathbb Z_k$ Zamolodchikov-Fateev parafermions \cite{ZF} 
with bosonic fields as their fractional superpartners. }
Existence of new string-theory dual descriptions in non-critical string theory is especially attractive 
because this would provide a new experimental laboratory to extract non-trivial information
about string theory. Therefore, it would be interesting to ask 
{\em what kind of phenomena appear beyond this infinite variety of the correspondence?}. 
In this paper, as a first step toward this study, 
we quantitatively analyze the multi-cut two-matrix models and their geometries
in the fractional-superstring critical points. 

As is reviewed in section \ref{FScriticalpointsSection}, 
the fractional-superstring critical points are special critical points in the $k$-cut two-matrix models. 
The multi-cut matrix models are generally shown \cite{fi1} to be controlled by multi-component KP hierarchy 
\cite{kcKP}. 
The fractional-superstring critical points are characterized by 
the following form of the $k$-component KP Lax pairs $(\bP,\bQ)$ \cite{irie2}:
\begin{align}
\bP(t,\del) = 
\Gamma \del^{\hat p}+ \sum_{n=0}^{\hat p-1} H_n(t) \del^{n},\qquad 
\bQ(t,\del) = 
\Gamma \del^{\hat q}+ \sum_{n=0}^{\hat q-1} \widetilde H_n(t) \del^{n},
\end{align}
with the two different choices of the leading matrix $\Gamma$ \cite{CISY1}: 
\begin{align}
\Gamma \equiv 
\begin{pmatrix}
0 & 1 & \cr
 & \ddots   & \ddots \cr
 &    &      0     & 1 \cr
1&  &             & 0
\end{pmatrix} 
\qquad \text{or}\qquad \Gamma \to \Gamma^{(\rm real)}\equiv 
\begin{pmatrix}
0 & 1 & \cr
 & \ddots   & \ddots \cr
 &    &      0     & 1 \cr
-1&  &             & 0
\end{pmatrix}.  \label{GammaIntroduction}
\end{align}
These choices of the leading matrices generally break 
the intrinsic $\mathbb Z_k$ symmetry of the multi-cut two-matrix models ($X$ and $Y$ are the two matrices),%
\footnote{Note that, in our analysis, the two-cut cases still preserve the $\mathbb Z_2$ symmetry. }
\begin{align}
(X,Y) \mapsto (\omega^n X ,\omega^{-n} Y)\qquad (\omega \equiv  e^{2\pi i /k}),  \label{ZkChargeConjInt}
\end{align}
and the system requires rather non-trivial analysis as compared 
with the $\mathbb Z_k$ symmetric critical points \cite{CISY1}. 
Despite of this fact, we show that these systems turn out to be still solvable even 
in the presence of {\em an arbitrary number of cuts}. 

The actual quantities investigated here are {\em macroscopic loop amplitudes} 
\cite{BIPZ,KazakovSeries,Kostov1,Kostov2,BDSS,Kostov3,MSS,DKK,Kris,AnazawaIshikawaItoyama1,AnazawaIshikawaItoyama2,AnazawaItoyama},%
\footnote{In this section, in order to avoid complexity of the notation, 
we do not distinguish the parameter $x$ and its scaling parameter $\zeta$, although they are distinguished 
in the main text. }
\begin{align}
Q(x) \sim \frac{1}{N} \vev{\tr \frac{1}{x-X}} 
= \int dXdY\, e^{-N\tr w(X,Y)} \Bigl[\frac{1}{N} \tr \frac{1}{x-X}\Bigr],
\end{align}
and they are known to play important roles in extracting various information about the matrix models: 
eigenvalue distributions \cite{BIPZ}, generating functions of local (on-shell) operators on worldsheets, 
and also effective potentials of a single (pair of) matrix-model eigenvalues $(x,y)$ \cite{David,KazakovKostov}. 
Furthermore, since these amplitudes directly correspond to boundary states of FZZT branes \cite{FZZT}, 
it is through the study of these amplitudes that we can perform a direct comparison 
with the amplitudes in the Liouville theory \cite{SeSh}. 
Since the worldsheet CFT calculation of fractional super Liouville field theory \cite{irie2} 
is far from completion, the study of macroscopic loop amplitudes is 
important not only for an understanding of the actual dynamics of this system, 
but also it could provide an important clue to the analysis of this not-yet solved Liouville theory. 

In this paper, the macroscopic loop amplitudes 
are studied within the Daul-Kazakov-Kostov prescription \cite{DKK} 
and its multi-cut extension \cite{CISY1}. In the DKK prescription 
(reviewed in the begining of section \ref{SectionCoshSinh}), 
the macroscopic loop amplitude $Q(x)$ is identified with the eigenvalues of the Lax operator $\bQ(t;\del)$. 
It turns out that there are generally two kinds of solutions 
in the fractional-superstring critical points: 
One solution is given by hyperbolic cosine and the other is by hyperbolic sine functions: 
\begin{align}
\text{cosh solutions:}&\qquad 
x = \sqrt{\mu} \cosh(\hat p \tau + 2\pi i \nu_j),\qquad 
Q^{(j)} = \mu^{\frac{\hat q}{2\hat p}} \cosh(\hat q \tau + 2\pi i \nu_j), \nn\\
\text{sinh solutions:}&\qquad 
x = \sqrt{\mu} \sinh(\hat p \tau + 2\pi i \nu_j),\qquad 
Q^{(j)} = \mu^{\frac{\hat q}{2\hat p}} \sinh(\hat q \tau + 2\pi i \nu_j), \label{CSsolIntroduction}
\end{align}
in the background with the bulk cosmological constant $\mu$. 
The integer $j=1,2,\cdots,k$ is the label of eigenvalues of the Lax operator $\bQ(t;\del)$. 
The sinh solution only appear 
when $k$ is even, and the cosh solutions appear generally. 
Therefore, the odd $k$ cases are natural extension of the bosonic string amplitudes \cite{Kostov1}; 
the even $k$ cases are natural extension of type 0 superstring amplitudes \cite{SeSh}. 
Interestingly, these general formulae naturally include bosonic $(k=1)$ and superstring $(k=2)$ cases, 
and the properties of algebraic curves and their singular points constitute natural generalizations of these 
special cases. This continuous extrapolation makes our current study distinct 
from the $\mathbb Z_k$ symmetric critical points \cite{CISY1}. 

The phase shift $\nu_j$ in the solutions in Eq.~\eq{CSsolIntroduction} is given by some proper rational number, 
and only this part of the amplitudes depends on the models we consider: 
the number of cuts, $k$, and the hermiticity of the two-matrix models, i.e. the leading Gamma matrices 
\eq{GammaIntroduction}:
\begin{align}
\nu_j = 
\left\{
\begin{array}{ll}
\ds
\frac{(j-1)}{k} & : \text{$\omega^{1/2}$-rotated potentials \ $(\Leftrightarrow \Gamma)$}, \cr
\ds
\frac{(2j-1)}{2k} & : \text{real potentials \ $(\Leftrightarrow \Gamma^{(\rm real)})$},
\end{array}
\right.
\end{align}
with $j=1,2,\cdots, k$. 
The first cases are realized in $\omega^{1/2}$-rotated (complex) potentials and the second are in real potentials. 
These two kinds of critical points are different when $k\in 4\mathbb Z$. These are the quantitative results 
obtained in this paper which enable us to extract the algebraic geometry realized in 
the weak-coupling regime of the multi-cut two-matrix models/fractional superstring theories. 

In addition to these quantitative results, we also discuss several implications based on our analysis 
as follows:%
\footnote{Further checks of these considerations 
should be of importance in future {\em non-perturbative} investigations. }

From the first sight, it might seem strange that the single macroscopic loop
amplitude $Q(x)$ or resolvent operator corresponds to the $k$ different amplitudes $Q^{(j)}(x)$ 
($j=1,2,\cdots,k$) which are realized as eigenvalues of the Lax operator $\bQ(t;\del)$. 
However we argue in section \ref{MultiCutGeometryStokesSection} 
that they play an important role to make up the {\em $k$-cut geometry} anticipated 
from the definition of the $k$-cut two-matrix models. 
The idea is based on the study about {\em Stokes phenomena} in the Baker-Akhiezer functions \cite{MMSS,SeSh2}. 
That is, although each amplitude $Q^{(j)}(x)$ has maximally {\em two cuts} on their branch, 
the resolvent operator $Q(x)$ can have $k$ semi-infinite cuts by being patched with $k$ different 
amplitudes $Q^{(j)}(x)$ in {\em different Stokes sectors} of the multi-cut Baker-Akhiezer function. 
The patching rule is also proposed in section \ref{MultiCutGeometryStokesSection}. 

\begin{figure}[htbp]
 \begin{center}
  \includegraphics[scale=0.75]{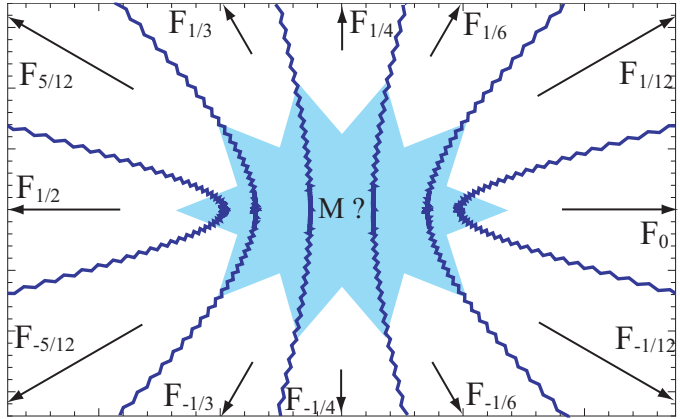}
 \end{center}
 \caption{\footnotesize 
A typical geometry of 12-cut fractional-superstring critical points ($x$ space). 
Each asymptotic regime around the arrow 
(denoted as $F_\nu$) is a perturbatively isolated 
sector, which corresponds to the irreducible algebraic curve $F^{(\nu)}(x,Q)=0$. 
This string theory includes six perturbative string theories which are also realized in 
$k$-fractional superstring theory of $k=1,2,3,6,12$. 
In particular, $F_0,F_{1/2}$ are identified with bosonic strings, 
and $F_{\pm 1/4}$ are with (two-cut phase of) type 0 superstring theory. 
Around the center is the strong coupling regime. } 
 \label{12cuts}
\end{figure}

The reduction of the number of cuts on each branch is also related to the highly reducibility 
of our resulting algebraic curve of the macroscopic loop amplitude. That is, we show that
the algebraic curve is factorized into $\lfloor \frac{k}{2}\rfloor$ irreducible curves:%
\footnote{In this paper, the maximal integer less than a real number $a$ is written as $\lfloor a \rfloor$, 
e.g. $\lfloor 1/2 \rfloor =0$ and $\lfloor -1/2 \rfloor =-1$. }
\begin{align}
F(x,Q) = \prod_{j=1}^{\lfloor \frac{k}{2}\rfloor} F^{(\nu_j)}(x,Q)=0. 
\end{align}
Each irreducible curve corresponds to a pair of eigenvalues of $\bQ(t;\del)$, 
\begin{align}
F^{(\nu_j)}(x,Q)= 0 \qquad \Leftrightarrow \qquad 
Q^{(j)}(x) \quad\text{and}\quad Q^{(k-j+2)}(x). 
\end{align}
As a general belief in matrix models,
if the algebraic curve is factorized into irreducible pieces, 
each irreducible sector does not communicate with other sectors in all order perturbation theory.%
\footnote{One of the intuitive understanding is following: Local operators are related to 
deformation of moduli of the curve. Connections between factorized curves are related to 
singular points which can open and connect the curves only by instanton effects. 
The correlators are 
determined by integrability of the deformation. However, deformations of each factorized curve are
independent on the deformations of other curves. Therefore, correlators between factorized curves 
are generally zero in the all-order perturbation theory. This kind of argument can be found in 
\cite{SeSh,KOPSS,fim}. \label{PerturbativeIsolation}} 
Therefore, these considerations indicate that in the spacetime $x$ of $(\hat p,\hat q)$ 
minimal $k$-fractional superstring theory,%
\footnote{See the spacetime interpretation of $x$ which has been given by \cite{fy3} \cite{MMSS,SeSh2}. }
or in the weak-coupling {\em Landscape} of this fractional superstring theory, 
there are $\lfloor \frac{k}{2}\rfloor$ perturbatively isolated vacua labeled by 
$j=1,2,\cdots,\lfloor \frac{k}{2}\rfloor$. Therefore, this non-perturbative string theory 
includes various perturbative string theories like its super-selection sectors, 
although these sectors can interact with each other in the non-perturbative formulation.%
\footnote{This string theory has a unique non-perturbative vacuum as a superposition of these perturbative vacua
because the orthonormal polynomials are unique if one fixes the bounded matrix-model potential. } 
A typical landscape of minimal $12$-fractional superstring theory is shown in Fig.~\ref{12cuts}. 

Furthermore, if one takes $k\to \infty$,%
\footnote{We would like to thank Ivan Kostov for drawing our attention to this limit, 
which eventually led us to this intriguing interpretation of the multi-cut matrix models. }
the theory becomes $(\hat p,\hat q)$ minimal $\infty$-fractional superstring theory 
and this theory includes all the perturbative vacua realized in 
$(\hat p,\hat q)$ minimal $k$-fractional superstring theory (with arbitrary number of $k=1,2,\cdots$, 
and $(\hat p,\hat q)$ is fixed). 
In this sense, $\infty$-fractional superstring theory can be viewed as a mother theory of 
perturbative fractional superstring theory. 

Interestingly in this limit, the labeling of perturbative vacua $j$ or $\nu_j$ 
forms a continuum $U(1)$ {\em angular} direction
which can only be observed 
by non-perturbative correction of string theory. Since the $\mathbb Z_k$ charge conjugation of 
D-branes is obtained by a translation of this angular coordinate $\nu \to \nu + a$, 
the fractional superstring theory can be understood as a Kaluza-Klein reduction 
along this non-perturbative direction $\nu$ in a three-dimensional gravity theory. 
This situation resembles the {\em non-critical M theory} proposed by 
P.~Ho\v rava and C.~A.~Keeler \cite{NonCriticalMTheory}
 as an extension of $\hat c=1$ type 0A/0B superstring theories. 
A possible connection to non-critical M theory is also disccused in section \ref{SecConclusion}. 

Organization of this paper is as follows: 
In section \ref{MacroscopicLoopAmplitudesSection}, 
after the basics of the fractional-superstring critical points are reviewed in section \ref{FScriticalpointsSection}, 
macroscopic loop amplitudes in these critical points are investigated in section \ref{SectionCoshSinh}. 
Section \ref{ComplexPotentialSection} is for the $\omega^{1/2}$-rotated-potential critical points, 
and Section \ref{RealPotentialModels} is for the real-potential critical points and 
the comparison between $\omega^{1/2}$-rotated-potential critical points 
and real-potential critical points is shown. In Section \ref{MatrixRealizationSection}, 
the actual expression of the Lax operators for the solutions is shown. 

In section \ref{AlgebraicStructuresSection}, 
the algebraic properties are studied. 
The algebraic equations are derived in section \ref{AlgebraicEquationsSection}, 
branch points are studied in section \ref{BranchCurvesSection}
and singular points are studied in section \ref{AlgSingularPointsSection}. 
In section \ref{MultiCutGeometryStokesSection}, the multi-cut geometry of macroscopic loop amplitudes
is discussed. In section \ref{SectionFZZTandZZamp}, 
the FZZT- and ZZ-brane amplitudes are shown. 
Section \ref{SecConclusion} is devoted to discussion, which includes a possible connection to 
non-critical M theory. 

\section{Macroscopic loop amplitudes \label{MacroscopicLoopAmplitudesSection}}
\subsection{The fractional-superstring critical points \label{FScriticalpointsSection}}

Here we review some basic facts about the multi-cut two-matrix models in the fractional-superstring critical points. 
One can find more details and references on fractional superstring theory in \cite{irie2} 
and on the multi-cut matrix models in \cite{CISY1}. 

The $k$-cut two-matrix models 
\begin{align}
\mathcal Z = \int_{M_N(\mathcal C_k) \times M_N(\mathcal C_k)} d X d Y e^{-N \tr w(X,Y)}, \label{DefinitionOfTwoMat}
\end{align}
are an extension of the usual two-matrix models \cite{Mehta}, and
defined by integrating over the set of $N\times N$ normal matrices 
$M_N(\mathcal C_k) \times M_N(\mathcal C_k)$ whose eigenvalues are along the 
$\mathbb Z_k$ symmetric radial lines $\mathcal C_k$:
\begin{align}
\mathcal C_k \equiv \bigcup_{l=1}^k \mathbb R \, \omega^l \subset \mathbb C\qquad (\omega\equiv e^{2\pi i /k}). 
\label{MContour}
\end{align}
The two-matrix potentials $w(x,y) \,(\equiv V_1(x)+V_2(y) -\theta xy)$ 
are generally chosen to be polynomials and there are two possible definitions of hermiticity:%
\footnote{This kind of complex potentials was first introduced by \cite{HMPN} in the 
two-cut one-matrix models, 
and then extended to the multi-cut two-matrix models by \cite{CISY1}.}
\begin{align}
\text{$\omega^{1/2}$-rotated potentials:}&\qquad 
w^*(X,Y) = w(\omega  X,\omega^{-1}Y), \nn\\
\text{real potentials:}&\qquad 
w^*(X,Y) = w(X,Y). 
\label{HermiticityPotentials}
\end{align}
Here $w^*(x,y)$ means complex conjugation of the function $w(x,y)$.%
\footnote{In our notation, $(w(x,y))^* = w^* (x^*,y^*)$.  }
These two definitions of hermiticities are the same if the potentials are $\mathbb Z_k$ symmetric: 
$w(\omega X,\omega^{-1}Y) = w(X,Y)$; 
Otherwise, these two definitions will give rise to different 
kinds of potentials. Each kind of potentials is shown to admit critical points, i.e. critical potentials \cite{CISY1},%
\footnote{The systematic study of the critical points in the two-matrix models was first given in 
\cite{TadaYamaguchiDouglas}\cite{DKK} and extended to the multi-cut cases in
\cite{CISY1}. }
and these critical potentials of different kinds can be related by analytic continuation of 
a parameter in the potentials: For example, the $(\hat p,\hat q,k)=(2,3;3)$ 
fractional-superstring critical potential \cite{CISY1} is given as 
\begin{align}
w_{\rm crit}^{(\omega^{1/2})}(x,y) = w_{\rm crit}(x,y;\omega^{-1} g),\qquad
w_{\rm crit}^{(\rm real)}(x,y) =  w_{\rm crit}(x,y; g), 
\end{align}
with a non-zero real number, $g \in \mathbb R\setminus \{0\}$, and the potentials, 
$w_{\rm crit}(x,y; g) \equiv V_1^{\rm crit}(x;g)+V_2^{\rm crit}(y;g) - \theta xy$, is given by
\begin{align}
\left\{
\begin{array}{l}
\ds
V_1^{\rm crit}(x;g) = 3g x^2 + \frac{13 x^3}{3} - \frac{12 g x^5}{5}- \frac{13 x^6}{6}+ \frac{g x^8}{8} + \frac{x^9}{9}, \nn\\
\ds
V_2^{\rm crit}(y;g) = gy + \frac{5 g^2 y^2}{2}- \frac{13 y^3}{3}- \frac{5 gy^4}{4}+ \frac{y^6}{6},
\end{array}
\right. \label{CriticalPotentials233}
\end{align}
with $\theta =26$.%
\footnote{With this particular choice of $\theta$, the parameter $R_*$ 
in Eq.~\eq{FractionalSuperstringCriticalPoints} becomes $R_*=1$. 
See \cite{CISY1} for detail.  }
That is, they are related by choosing a proper complex phase of each term in the potentials.%
\footnote{However we will see that the off-critical amplitudes are distinct 
and not related by simple transformation. }

Critical points of each case are characterised 
by the recursive relation of the orthonormal polynomials,
\begin{align}
&x \alpha_n(x) = \sum_{s \in \mathbb Z} A_s(n) \, e^{s \del_n}\alpha_{n}(x),
\qquad N^{-1} \frac{\del}{\del x} \alpha_n(x) 
= \sum_{s \in \mathbb Z} B_s(n) \, e^{s \del_n}\alpha_{n}(x), \nn\\
&y \beta_n(y) = \sum_{s \in \mathbb Z} C_s(n) \, e^{s \del_n}\beta_{n}(y),
\qquad N^{-1} \frac{\del}{\del y} \beta_n(y) 
= \sum_{s \in \mathbb Z} D_s(n) \, e^{s \del_n}\beta_{n}(y), 
\end{align}
with $\ds \int dx dy\, e^{-N w(x,y)}\, \alpha_m(x) \, \beta_n(y) = \delta_{m,n}$. 
This can be translated in terms of the Baker-Akhiezer functions 
of $k$-component KP hierarchy%
\footnote{The relationship between {\em multi}-cut matrix models 
and {\em multi}-component KP hierarchy 
was first pointed out in \cite{fi1}. } as follows:
\begin{align}
\zeta \, \psi (t;\zeta ) = \mathcal A (t;\del) \, \psi (t;\zeta ),
\qquad g_{\rm str} \frac{\del }{\del \zeta } \, \psi (t;\zeta) 
= \mathcal B (t;\del) \,\psi (t;\zeta), \nn\\
\tilde \zeta \, \widetilde\psi (t;\tilde \zeta ) 
= \mathcal C (t;\del) \, \widetilde \psi (t;\tilde \zeta ),
\qquad g_{\rm str} \frac{\del }{\del \tilde \zeta } \, \widetilde \psi (t;\tilde \zeta) 
= \mathcal D (t;\del) \,\widetilde \psi (t;\tilde \zeta). \label{SmoothOrthPoly}
\end{align}
Here $\psi(t;\zeta)$ is a $k$-vector Baker-Akhiezer function which can be derived%
\footnote{The relation between $\psi(t;\zeta)$ and $\alpha_n(x)$ was first proposed by
\cite{MultiCut} in one-matrix models. The case of the multi-cut two-matrix models 
is given in \cite{CISY1}. }
from the orthonormal polynomials $\alpha_n(x)$, 
and $\widetilde \psi(t;\zeta)$ is its dual which comes from the dual polynomials $\beta_n(y)$. 
The scaling parameters are defined as
\begin{align}
&N^{-1} = g_{\rm str} \,a^{\frac{\hat p+\hat q}{2}} \to 0,\qquad 
\frac{n}{N} = \exp\bigl(-t a^{\frac{\hat p+\hat q-1}{2}}\bigr) \to 1,\nn\\
&\del_n = - a^{1/2} \,g_{\rm str}\, \del_t \equiv - a^{1/2}\del  \to 0, 
\qquad(a\to 0). 
\end{align} 
and the boundary cosmological constants $\zeta$ as
\begin{align}
\text{$\omega^{1/2}$-rotated potentials:}&\qquad
x = \omega^{1/2} a^{\frac{\hat p}{2}} \zeta \to 0,\qquad y = \omega^{-1/2} a^{\frac{\hat q}{2}} \tilde \zeta \to 0, \nn\\
\text{real potentials:}&\qquad
x = a^{\frac{\hat p}{2}} \zeta \to 0,\qquad y = a^{\frac{\hat q}{2}} \tilde \zeta \to 0. \label{RelXandZeta}
\end{align}
The Lax operators in Eq.~\eq{SmoothOrthPoly} are related by 
\begin{align}
\text{$\omega^{1/2}$-rotated potentials:}&\qquad
\left\{
\begin{array}{l}
\mathcal B = \omega^{1/2}  a^{-\frac{\hat q}{2}} V_1'( \omega^{1/2} a^{\frac{\hat p}{2}}\mathcal A) 
- \mathcal C^{\rm T},\cr
\mathcal D = \omega^{-1/2} a^{-\frac{\hat p}{2}} V_2'(\omega^{-1/2} a^{\frac{\hat q}{2}}\mathcal C) 
- \mathcal A^{\rm T}, 
\end{array}
\right. \nn\\
\text{real potentials:}&\qquad
\left\{
\begin{array}{l}
\mathcal B = a^{-\frac{\hat q}{2}} V_1'(a^{\frac{\hat p}{2}}\mathcal A) - \mathcal C^{\rm T},\cr
\mathcal D = a^{-\frac{\hat p}{2}} V_2'(a^{\frac{\hat q}{2}}\mathcal C) - \mathcal A^{\rm T},
\end{array}
\right. 
\end{align}
and in particular, in the fractional-superstring critical points $(\hat p,\hat q;k)$ \cite{irie2}, 
the pair $(\mathcal A, \mathcal C)$ 
is shown to be expressed as follows \cite{CISY1}:
\begin{align}
\text{$\omega^{1/2}$-rotated potentials:}\quad 
\left\{
\begin{array}{l}
\ds
\mathcal A(t;\del) = \sqrt{R_*} (-1)^{\hat p}\times  \Gamma \,\del^{\hat p} 
+ \sum_{n=1}^{\hat p} H_n(t)\, \del^{\hat p-n},  \cr
\ds
\mathcal C^{\rm T}(t;\del) = C \sqrt{R_*} k^{\hat q}\times  \Gamma \, \del^{\hat q} 
+ \sum_{n=1}^{\hat q} \widetilde H_n(t)\, \del^{\hat q-n},
\end{array}
\right. \label{FractionalSuperstringCriticalPoints}
\end{align}
with the matrix $\Gamma$ which is given by the $k\times k$ {\em shift matrix}, 
\begin{align}
\Gamma \equiv
\begin{pmatrix}
0 & 1 & \cr
 & \ddots   & \ddots \cr
 &    &      0     & 1 \cr
1&  &             & 0
\end{pmatrix},\qquad \Gamma^k=I_k.
\end{align}
The matrix $I_k$ is the $k\times k$ unit matrix. $R$ and $C$ are non-zero real constants.%
\footnote{$R$ is a critical value of $R_n=h_n/h_{n-1}$ with $\alpha_n(x) = x^n /\sqrt{h_n} +\cdots,\ 
\beta_n(y) = y^n/\sqrt{h_n} +\cdots$. See the definition in \cite{CISY1}.}
An important consequence from the matrix models is that {\em all the coefficient 
functions $H_n(t)$ and $\widetilde H_n(t)$ in Eq.~\eq{FractionalSuperstringCriticalPoints} 
are real functions} \cite{CISY1}.
In the case of real-potential critical points, what we need to do is 
just replacing the shift matrix $\Gamma$ by $\Gamma^{\rm (real)}$: 
\begin{align}
\Gamma \to \Gamma^{\rm (real)} \equiv 
\begin{pmatrix}
0 & 1 & \cr
 & \ddots   & \ddots \cr
 &    &      0     & 1 \cr
-1&  &             & 0
\end{pmatrix},\qquad \bigl(\Gamma^{(\rm real)}\bigr)^k = - I_k, 
\end{align}
that is, 
\begin{align}
\text{real potentials:}\quad 
\left\{
\begin{array}{l}
\ds
\mathcal A(t;\del) = \sqrt{R_*} (-1)^{\hat p}\times  \Gamma^{(\rm real)} \,\del^{\hat p} 
+ \sum_{n=1}^{\hat p} H_n^{(\rm real)}(t)\, \del^{\hat p-n},  \cr
\ds
\mathcal C^{\rm T}(t;\del) = C \sqrt{R_*} k^{\hat q}\times  \Gamma^{(\rm real)} \, \del^{\hat q} 
+ \sum_{n=1}^{\hat q} \widetilde H_n^{(\rm real)}(t)\, \del^{\hat q-n}. 
\end{array}
\right. \label{FractionalSuperstringCriticalPointsReal}
\end{align}
Here all the coefficient functions $H_n^{(\rm real)}(t)$ and $\widetilde H_n^{(\rm real)}(t)$ 
in Eq.~\eq{FractionalSuperstringCriticalPointsReal} 
are also real functions \cite{CISY1}.

From these Lax operators, one can conclude that the fractional-superstring 
critical points (except for two-cut cases) do not have the $\mathbb Z_k$ symmetry even right at the critical points, 
\begin{align}
w_{\rm crit}(\omega X,\omega^{-1}Y;g) \neq w_{\rm crit}(X,Y;g), 
\end{align}
since the $\mathbb Z_k$ symmetry requires that the leading matrix of operator $\mathcal C$
should be $\Gamma^{-1}$ \cite{irie2,CISY1}. 
Note that the critical points of even-number-cut models admit the following residual $\mathbb Z_2$ symmetry:
\begin{align}
w_{\rm crit}(- X,-Y;g) = w_{\rm crit}(X,Y;g),  \label{residualZ2Sym}
\end{align}
and the Lax operators $\mathcal X \, (=\mathcal A,\mathcal B,\mathcal C,\mathcal D)$ 
satisfy the following constraint:
\begin{align}
\{E,\mathcal X\} = 0,
\end{align}
with the $\mathbb Z_2$ grading matrix $E$ which is defined by 
\begin{align}
E \equiv \underset{\frac{k}{2}}{\underbrace{\sigma_3 \oplus \sigma_3 \oplus \cdots \oplus \sigma_3}} =
\begin{pmatrix}
1 & \cr
& -1 & \cr
&     & \ddots & \cr
& & &1 & \cr
& & && -1  
\end{pmatrix}. 
\end{align}
In particular, the two-cut ($k=2$) 
critical potentials preserve the $\mathbb Z_2$ symmetry at the critical points.%
\footnote{One can also say that it is because the gamma matrices satisfy $\Gamma= \Gamma^{-1}$ 
in the two-cut cases. In this sense, the $\mathbb Z_2$-breaking critical points in the two-cut matrix models
are missing in our analysis. Since they correspond to string theories flowed by Ramond-Ramond perturbation
of the type 0 superstring theory \cite{UniCom}, it is interesting system to study. }

In the section below, we consider the following special pair of the Lax operators:
\begin{align}
\bP (t;\del) \equiv \mathcal A(t;\del),\qquad \bQ (t;\del) \equiv - \mathcal C^{\rm T}(t;\del)\, 
\Bigl(= \mathcal B - a^{-\frac{\hat q}{2}} V_1'(a^{\frac{\hat p}{2}}\mathcal A)\Bigr), \label{PQACrel1}
\end{align}
which satisfies the Douglas equation \cite{DouglasGeneralizedKdV}: $[\bP,\bQ]= g_{\rm str} I_k$. 
Note that the operators on the dual side $(\widetilde \bP,\widetilde \bQ)$ are simply related by transpose:
\begin{align}
\widetilde \bP (t;\del) =-\bQ^{\rm T} (t;\del) \equiv \mathcal C(t;\del), 
\qquad \widetilde \bQ (t;\del) = -\bP^{\rm T}(t;\del) \equiv - \mathcal A^{\rm T}(t;\del), \label{PQACrel2}
\end{align}
which also satisfy the Douglas equation: $[\widetilde \bP,\widetilde \bQ]= g_{\rm str} I_k$. 
Consequently, the recursive relation of orthonormal polynomials \eq{SmoothOrthPoly} can be 
reexpressed as
\begin{align}
\zeta \, \Psi (t;\zeta ) = \bP (t;\del) \, \Psi (t;\zeta ),
\qquad g_{\rm str} \frac{\del }{\del \zeta } \, \Psi (t;\zeta) 
= \bQ (t;\del) \,\Psi (t;\zeta), \nn\\
\tilde \zeta \, \widetilde\Psi (t;\tilde \zeta ) 
= \widetilde \bP (t;\del) \, \widetilde \Psi (t;\tilde \zeta ),
\qquad g_{\rm str} \frac{\del }{\del \tilde \zeta } \, \widetilde \Psi (t;\tilde \zeta) 
= \widetilde \bQ (t;\del) \,\widetilde \Psi (t;\tilde \zeta), \label{SmoothOrthPoly2}
\end{align}
with
\begin{align}
\Psi(t;\zeta) 
= \psi(t;\zeta) \,
\exp\Bigl[
\frac{-a^{-\frac{\hat p+\hat q}{2}}}{g_{\rm str}} V_1\bigl(a^{\frac{\hat p}{2}}\zeta\bigr) 
\Bigr],\quad
\widetilde \Psi(t;\tilde \zeta) 
= \widetilde \psi(t;\tilde \zeta) \,
\exp\Bigl[
\frac{-a^{-\frac{\hat p+\hat q}{2}}}{g_{\rm str}} V_2\bigl(a^{\frac{\hat q}{2}}\tilde \zeta\bigr) 
\Bigr],
\end{align}
and $\psi(t;\zeta)$ and $\widetilde \psi(t;\tilde \zeta)$ are defined in Eq.~\eq{SmoothOrthPoly}. 

\subsection{The cosh and sinh solutions \label{SectionCoshSinh}}
In this section, macroscopic loop amplitudes are studied in the fractional-superstring critical points. 
We adopt the Daul-Kazakov-Kostov prescription \cite{DKK} and its multi-cut generalization \cite{CISY1} 
to obtain macroscopic loop amplitudes. 
First we note the relationship between macroscopic loop amplitudes and the Lax operators 
\cite{DKK}\cite{Moore,MMSS}\cite{fim}\cite{fi1}. The macroscopic loop amplitudes we consider 
are following:
\begin{align}
\del_x\Omega (x)\equiv \vev{\frac{1}{N}\tr \frac{1}{x-X}},&\qquad 
Q(\zeta)\equiv 
a^{-\frac{\hat q}{2}} \del_x\Omega (a^{\frac{\hat p}{2}}\zeta )
-a^{-\frac{\hat q}{2}} V_1' (a^{\frac{\hat p}{2}}\zeta ),\nn\\
\del_y\widetilde \Omega (y)\equiv \vev{\frac{1}{N}\tr \frac{1}{y-Y}},&\qquad
\widetilde Q(\tilde \zeta)\equiv
a^{-\frac{\hat p}{2}} \del_y\widetilde \Omega (a^{\frac{\hat q}{2}}\tilde \zeta )
- a^{-\frac{\hat p}{2}} V_2' (a^{\frac{\hat q}{2}}\tilde \zeta ).  \label{MacroResol}
\end{align}
The basic idea comes from the exact expression of orthonormal polynomials \cite{GrossMigdal2},
\begin{align}
\alpha_n(x) = \frac{1}{\sqrt{h_n}}\vev{\det \bigl(x-X_{(n)}\bigr)}_{n\times n},
\qquad \beta_n(y) = \frac{1}{\sqrt{h_n}}\vev{\det \bigl(y-Y_{(n)}\bigr)}_{n\times n}.  \label{OrthAlphaBeta}
\end{align}
Here $\vev{\cdots}_{n\times n}$ is the matrix integral of the $n\times n$ truncated matrices 
$(X_{(n)},Y_{(n)})$, 
\begin{align}
\vev{\det \bigl(x-X_{(n)}\bigr)}_{n\times n} 
\equiv \int_{M_n(\mathcal C_k) \times M_n(\mathcal C_k)} dX_{(n)}d Y_{(n)}\, e^{-N \tr w(X_{(n)},Y_{(n)})} 
\det \bigl(x-X_{(n)}\bigr). 
\end{align}
From these relations, Eqs.~\eq{MacroResol} and \eq{OrthAlphaBeta}, we expect that 
the eigenvalues of the Lax operators \eq{SmoothOrthPoly} 
are related to the macroscopic loop amplitudes in the following way:
\begin{align}
N^{-1}\,\frac{\del}{\del x} \alpha_n(x) \sim \del_x \Omega(x)\, \alpha_n(x),
\qquad N^{-1}\,\frac{\del}{\del y} \beta_n(y) \sim \del_y \widetilde  \Omega(y)\, \beta_n(y),
\end{align}
in the large $N$ limit. 
This claim has been argued and proved in various contexts \cite{DKK} \cite{Moore,MMSS} \cite{fim}\cite{fi1}. 
In terms of  $(\bP,\bQ)$ of Eqs.~\eq{PQACrel1} and \eq{PQACrel2}, this means that these 
amplitudes can be
derived as the simultanious eigenvalues of the Lax pair:
\begin{align}
Q(\zeta):& \qquad (\zeta,Q) 
\sim \bigl(\bP(t;\del), \bQ(t;\del)\bigr),\nn\\
\widetilde Q(\tilde \zeta):& \qquad (\tilde \zeta,\widetilde Q) 
\sim \bigl(-\bQ^{\rm T}(t;\del),-\bP^{\rm T}(t;\del) \bigr), \label{MacroLaxRelation}
\end{align}
in the weak string coupling limit, $g_{\rm str}\to 0$, 
(or the dispersionless KP hierarchy limit). 
This problem was solved in the one-cut two-matrix models \cite{DKK}
and also in $\mathbb Z_k$ symmetric critical points of the multi-cut two-matrix models \cite{CISY1}. 
In the following, we study this prescription in the fractional-superstring critical points 
\eq{FractionalSuperstringCriticalPoints}. 
Although the correspondence seems to be between the single macroscopic loop amplitude $Q(\zeta)$ 
and the $k$ eigenvalues of the operator $\bQ$, this is argued to be consistent in section 
\ref{MultiCutGeometryStokesSection}. 

\subsubsection{The $\omega^{1/2}$-rotated-potential cases \label{ComplexPotentialSection}}
The system we consider is the Douglas equation $[\bP,\bQ]= g_{\rm str}I_k$ 
with the pair of $k\times k$ matrix valued differential operators $\bP(t;\del)$ 
and $\bQ(t;\del)$ defined in Eqs.~\eq{FractionalSuperstringCriticalPoints} and \eq{PQACrel1} 
in the $\omega^{1/2}$-rotated-potential critical points. The real-potential critical points are 
studied separately in section \ref{RealPotentialModels}. 
In the leading order of the week coupling expansion, one can factorize the operators 
into a dimensionful part and a dimensionless part without any ambiguity of ordering:%
\footnote{Here we drop the irrelevant coefficients of operators $\bP$ and $\bQ$ to simplify the discussion. }
\begin{align}
&\left\{
\begin{array}{l}
\ds
\bP (t;\del) = \lambda^{\hat p}\, \bPi(z) \equiv \lambda^{\hat p} \Bigl( \Gamma\, z^{\hat p} 
+ \sum_{n=1}^{\hat p} U_n \,z^{\hat p-n}\Bigr) + O(g_{str}), \cr
\ds
\bQ (t;\del) = \lambda^{\hat q}\, \bXi(z) \equiv \lambda^{\hat q} \Bigl( \Gamma\, z^{\hat q} 
+ \sum_{n=1}^{\hat q} V_n \,z^{\hat q-n} \Bigr) + O(g_{str}), 
\end{array}
\right. \label{DimensionSeparation}\\
&\qquad \text{with}\qquad \lambda \equiv t^{\frac{1}{\hat p+\hat q-1}},\qquad 
z= \lambda^{-(\hat p+\hat q)} g_{str}\, t\, \del_t.
\end{align}
Given these ansatz, one can show that the matrices satisfy the leading Douglas equation, 
\begin{align}
[\bP,\bQ] = g_{str} I_k 
\quad \Rightarrow \quad 
[\bPi(z),\bXi(z)] = 0 \qquad \text{: the leading equation},  \label{LeadingDouglasAA}
\end{align}
which means that the two matrix functions $\bPi(z)$ and $\bXi(z)$ commute with each other. 

In the next leading order of the Douglas equation, the ordering among $(z,\lambda)$ variables 
in the expression \eq{DimensionSeparation} could cause some problem \cite{CISY1}. 
This ordering problem can be solved by first solving simultanious diagonalization of operators 
$\bP(t;\del)$ and $\bQ(t;\del)$ in the week coupling limit,  
or equivalently $\bPi(z)$ and $\bXi(z)$:
\begin{align}
\bPi (z) \simeq& 
\begin{pmatrix}
\Pi^{(1)}_{\hat p}(z) & \cr
 & \Pi^{(2)}_{\hat p}(z) & \cr
 & & \ddots &\cr
 & & & \Pi^{(k)}_{\hat p}(z) 
\end{pmatrix} 
= 
\begin{pmatrix}
1 & \cr
 & \omega & \cr
 & & \ddots &\cr
 & & & \omega^{k-1} 
\end{pmatrix}
z^{\hat p} + O(z^{\hat p-1}),\nn\\
\bXi (z) \simeq& 
\begin{pmatrix}
\Xi^{(1)}_{\hat q}(z) & \cr
 & \Xi^{(2)}_{\hat q}(z) & \cr
 & & \ddots &\cr
 & & & \Xi^{(k)}_{\hat q}(z) 
\end{pmatrix}
= 
\begin{pmatrix}
1 & \cr
 & \omega & \cr
 & & \ddots &\cr
 & & & \omega^{k-1} 
\end{pmatrix}
z^{\hat q} + O(z^{\hat p-1}). \label{AsymPandQ}
\end{align}
Here $\simeq$ denotes equality upto some similarity transformation. 
The next leading order of the Douglas equation then turns out to be the EZJ-DKK equation 
\cite{EynardZinnJustin}\cite{DKK}, 
\begin{align}
\hat q \,\Pi^{(j)}_{\hat p}{}'(z) \, \Xi^{(j)}_{\hat q}(z) 
- \hat p\, \Pi^{(j)}_{\hat q}{}'(z) \, \Xi^{(j)}_{\hat p}(z) = \hat p+\hat q-1 
\qquad (j=1,2,\cdots,k). \label{EZJ-DKKeq}
\end{align}
Note that constant appearing on the right-hand side of Eq.~\eq{EZJ-DKKeq} 
is common for all $j=1,2,\cdots,k$.

Up to now the basic procedure to solve the equations is 
the same as the $\mathbb Z_k$ symmetric cases \cite{CISY1}. 
The major difference comes from the Lax operators \eq{DimensionSeparation} 
which break the $\mathbb Z_k$ symmetry; If there is the $\mathbb Z_k$ symmetry, 
the diagonalization is trivially solved \cite{CISY1}. 
Since there is no $\mathbb Z_k$ symmetry now, the characteristic equation of the Lax operators:
\begin{align}
\det\Bigl[\zeta I_k- \lambda^{\hat p} \bPi(z)\Bigr] =0, 
\qquad \det\Bigl[Q I_k- \lambda^{\hat q} \bXi(z)\Bigr] =0 \label{CharaPQ}
\end{align}
can have various kinds of solutions which are not algebraic in general. 
This makes it difficult to obtain general formulae of the eigenvalues in the fractional-superstring critical points.%
\footnote{It is possible to obtain complete solution of the diagonalization 
when the number of cuts, $k$, is smaller than 5, $k< 5$. }
In view of this, our strategy here is that we first propose a porper ansatz which can solve 
the EZJ-DKK equation \eq{EZJ-DKKeq}, 
and then show that they are actually the solutions to the characteristic equations \eq{CharaPQ}. 
In this procedure, the uniqueness of the solution is not granted but our purpose is to obtain 
non-trivial solutions of the system.%
\footnote{It is known that even in the one-cut cases, if one consider higher critical points $(p,q)$, 
there are several exceptional solutions which have no direct correspondence 
in the Liouville theory calculation \cite{DKK}. 
Therefore, it is also interesting to consider such a solution in our multi-cut system 
but it is out of the scope of this paper. }

Our ansatz for the $(\hat p,\hat q;k)$ fractional-superstring critical points is as follows: 
{\em the eigenvalues of $\bP$ and $\bQ$ are given by hyperbolic cosines 
with some proper phase shifts}, 
i.e., the eigenvalues $\Pi_{\hat p}^{(j)}(z)$ and $\Xi_{\hat q}^{(j)}(z)$ are written as
\begin{align}
\Pi^{(j)}_{\hat p}{}(z) 
=& e^{-2\pi i \delta}T_{\hat p}^{(\frac{j-1}{k}+\delta)}(z)
\equiv e^{-2\pi i \delta}\cosh\bigl(\hat p \tau + 2\pi i \frac{j-1}{k} + 2\pi i \delta\bigr), \nn\\
\Xi^{(j)}_{\hat q}{}(z) 
=& e^{-2\pi i \delta}T_{\hat q}^{(\frac{j-1}{k}+\delta)}(z)
\equiv e^{-2\pi i \delta}\cosh\bigl(\hat q \tau + 2\pi i \frac{j-1}{k} + 2\pi i \delta\bigr). 
\label{OurAnsatz}
\end{align}
Here we define {\em deformed Chebyshev functions},%
\footnote{Note that $T_n^{(\nu+m)}(z)= T_n^{(\nu)}(z)$ if $m\in \mathbb Z$. }
$T_n^{(\nu)}(\cosh\tau) \equiv  \cosh (n\tau + 2\pi i \nu)$, which can also be expressed as
\begin{align}
T_n^{(\nu)}(z) 
=& \cos(2\pi \nu)\, T_n(z) 
+i \sin(2\pi \nu)\, U_{n-1}(z) \sqrt{z^2-1} \nn\\
=& \frac{e^{2\pi i \nu} \Bigl(z+\sqrt{z^2-1}\Bigr)^n+e^{-2\pi i \nu} \Bigl(z-\sqrt{z^2-1}\Bigr)^n}{2} \nn\\
 \sim & 2^{n-1} e^{2\pi i \nu} z^{n}+\cdots \quad  (z \to \infty), \label{DeformedCheb}
\end{align}
where $z=\cosh \tau$. 
The polynomials $T_n(z)$ and $U_n(z)$ are the Chebyshev polynomials of the first and second kinds: 
$T_n(\cos \theta)= \cos(n\theta)$, $U_{n-1}(\cos \theta )= \sin(n\theta)/\sin \theta $. 
The phase $e^{-2\pi i \delta}$ in front of the hyperboic cosine of the solutions \eq{OurAnsatz} is necessary 
to have correct asymptotic behavior $z \to \infty$ (see Eq.~\eq{AsymPandQ}). 
Importantly, one can check that the deformed Chebyshev functions are always solutions 
to the EZJ-DKK equation%
\footnote{Although this is the equation for unitary series $(\hat p,\hat q)=(\hat p,\hat p+1)$, 
the EZJ-DKK equation for the general series $(\hat p,\hat q)$ is obtained by changing the canonical pair 
$(t,z)$ to $(\mu,w)$ in \cite{DKK}. Here $t$ is the most relevant coupling constant; 
$\mu$ is the cosmological constant and $w = T_{\hat q-\hat p}(z)$. 
Note that this extension is also applicable to the multi-cut cases:
\begin{align}
\hat qT_{\frac{\hat p}{\hat q-\hat p}}^{(\nu)}{}'(w)T_{\frac{\hat p}{\hat q-\hat p}}^{(\nu)}{}(w) 
-\hat p T_{\frac{\hat q}{\hat q-\hat p}}^{(\nu)}{}'(w)T_{\frac{\hat p}{\hat q-\hat p}}^{(\nu)}{}(w) 
= \frac{\hat p \hat q}{\hat q-\hat p}. 
\end{align}
Therefore, the defomed Chebyshev polynomials are general solutions to the EZJ-DKK equation. 
} 
with an arbitrary phase shift, $\nu \in \mathbb C$:
\begin{align}
(\hat p+1)T_{\hat p}^{(\nu)}{}'(z)T_{\hat p+1}^{(\nu)}{}(z) 
-\hat p T_{\hat p+1}^{(\nu)}{}'(z)T_{\hat p}^{(\nu)}{}(z) = \hat p(\hat p+1). 
\end{align}
Therefore, what we need to check next is which phase shift $\delta$ is allowed 
by the characteristic equations \eq{CharaPQ}. 

There are two constraints imposed by the characteristic equations \eq{CharaPQ}:
\begin{itemize}
\item Since the matrices $\bPi(z)$ and $\bXi(z)$ are matrix-valued polynomials in $z$, 
the characteristic equations of $\bP$ and $\bQ$ (Eq.~\eq{CharaPQ}) 
must be {\em polynomial equations} in $(\zeta,z)$ of degrees $(k,k\hat p)$ for $\bP$, 
and in $(Q,z)$ of degrees $(k,k\hat q)$ for $\bQ$. 
\item Since the coefficients of Lax operators $\bP$ and $\bQ$ are all real functions \cite{CISY1}, 
{\em all the coefficients of the characteristic equations \eq{CharaPQ} must be real functions.} 
\end{itemize}
Noticing that the characteristic equations of Eq.~\eq{OurAnsatz} are generally written as
\begin{align}
\prod_{j=1}^k 
\Bigl( \zeta - 
\lambda^{\hat p} e^{-2\pi i \delta}\, T_{\hat p}^{(\frac{j-1}{k}+ \delta)}(z)
\Bigr)= \frac{\lambda^{k\hat p} e^{2\pi i k\delta}}{2^{k-1}}\Bigl(
T_k(e^{2\pi i \delta}\zeta/ 
\lambda^{\hat p} ) - T_{k\hat p}^{(k\delta)}(z)
\Bigr),
\end{align}
one can see that the first requirement results in 
\begin{align}
k\delta = \frac{n}{2} \in \frac{1}{2}\mathbb Z,\qquad 
\text{($\Leftrightarrow$ $T_{k\hat p}^{(k\delta)}(z)$ is a polynomial. See Eq.~\eq{DeformedCheb}.)} 
\end{align}
and then the algebraic equation \eq{CharaPQ} is given by 
\begin{align}
\det\Bigl[\zeta I_k- \lambda^{\hat p} \bPi(z)\Bigr]
= \frac{\lambda^{k\hat p}}{2^{k-1}}\Bigl(
(-1)^n T_k(e^{\frac{\pi i n}{k}}\zeta/\lambda^{\hat p}) - T_{k\hat p}(z) \Bigr)=0. \label{CharaFirstSol} 
\end{align}
Furthermore, the second constraint on Eq.~\eq{CharaFirstSol} 
(i.e. $T_k(e^{\frac{\pi i n}{k}}\zeta/\lambda^{\hat p})$) 
requires that the phase shift must be
\begin{align}
\delta =\frac{n}{2k}\in \frac{\mathbb Z}{2}\qquad \text{if $k$ is odd},\qquad
\delta =\frac{n}{2k}\in \frac{\mathbb Z}{4}\qquad \text{if $k$ is even}. 
\end{align}
Therefore, the fractional-superstring critical points of the multi-cut matrix models 
admit two different kinds of solutions: 
\begin{align}
\underset{\ds \vspace{1cm}(\delta = 0,1/2)}{\underline{\text{cosh solution}}:}&\quad
\left\{
\begin{array}{l}
\ds 
\Pi^{(j)}_{\hat p}{}(z) = \cosh\bigl(\hat p \tau + 2\pi i \frac{j-1}{k}\bigr) 
= T_{\hat p}^{(\frac{j-1}{k})}(z), \cr
\ds 
\Xi^{(j)}_{\hat q}{}(z) = \cosh\bigl(\hat q \tau + 2\pi i \frac{j-1}{k}\bigr) 
= T_{\hat q}^{(\frac{j-1}{k})}(z), 
\end{array}
\right.  \label{CoshSolution} \\
\text{with}\quad &
\det \Bigl[ (2 \zeta ) \,I_k - \bP(z)\Bigr] 
= 2 \lambda^{k\hat p}\Bigl(T_k(\zeta/\lambda^{\hat p}) - T_{k\hat p}(z)\Bigr) =0, \nn\\
\underset{\ds \vspace{1cm}(\delta = \pm 1/4)}{\underline{\text{sinh solution}}:}&\quad
\left\{
\begin{array}{r}
\ds 
\Pi^{(j)}_{\hat p}{}(z) = \sinh\bigl(\hat p \tau + 2\pi i \frac{j-1}{k}\bigr) 
= i \,T_{\hat p}^{(\frac{j-1}{k}-\frac{1}{4})}(z), \cr
\ds 
\Xi^{(j)}_{\hat q}{}(z) = \sinh\bigl(\hat q \tau + 2\pi i \frac{j-1}{k}\bigr) 
= i \, T_{\hat q}^{(\frac{j-1}{k}-\frac{1}{4})}(z). 
\end{array}
\right.  \label{SinhSolution} \\
\text{with}\quad & \det \Bigl[ (2 \zeta ) \,I_k - \bP (z)\Bigr] 
= 2 \lambda^{k\hat p}
\Bigl( (-1)^{\frac{k}{2}}T_k(-i \zeta/\lambda^{\hat p}) - T_{k\hat p}(z)\Bigr) =0. \nn
\end{align}
These solutions are referred to as {\em cosh solutions} 
and {\em sinh solutions}, respectively. Existence of these solutions depends on the 
parity of $k$:
\begin{align}
\text{$k$ is odd}:&\qquad \text{cosh solutions}, \nn\\
\text{$k$ is even}:&\qquad \text{cosh solutions and sinh solutions}. 
\end{align}
Note that the sinh solutions are essentially equivalent to the cosh solutions 
when $k \in 4\mathbb Z$:
\begin{align}
\sinh\bigl(\hat p \tau + 2\pi i \frac{j-1}{k}\bigr) 
= i \cosh\bigl(\hat p \tau + 2\pi i \frac{j'-1}{k}\bigr),\qquad 
\bigl(j'=j-\frac{k}{4}\bigr).  \label{4nmidkSeqC}
\end{align}
Consequently, the cases of $k\in 4\mathbb Z$ can be understood as 
a mixture of bosonic and type 0 superstring amplitudes. 
In particular, the cases of $k = 4$ include the one-cut phase and two-cut phase 
solutions of type 0 superstrings (see \cite{UniCom,SeSh}) at the same time. In this sense, it tempts us 
to interpret this system as type 0 superstring theory with the $\eta=\pm1$ FZZT branes, 
but the annulus amplitudes among them \cite{Irie1} seem to be different.%
\footnote{
This can be understood as follows: 
The CFT calculation tells us that annulus amplitdes between $\eta=+1$ and $\eta=-1$ are 
non-local \cite{Irie1} {\em if $\eta=\pm1$ are treated equally}. However from the the free-fermion 
viewpoints \cite{fkn,fy12,fy3,fim,fi1}, the $\nu=0,1/2$ sector 
and the $\nu=\pm 1/4$ sector are treated equally \cite{fi1} and 
the annulus amplitudes between the $\nu=0,1/2$ sector 
and the $\nu=\pm 1/4$ sector cannot have non-local annulus amplitudes \cite{fim}. 
}
Also note that, on the CFT side, 
there is no ZZ brane which connects the $\eta=\pm$ FZZT branes \cite{SeSh}; 
in the case of $k\in 4\mathbb Z$, there are ZZ branes which connect the $\nu=0,1/2$ sector 
to the $\nu=\pm 1/4$ sector (see the next section).

\subsubsection{The real potentials v.s. $\omega^{1/2}$-rotated potentials 
\label{RealPotentialModels}}
The real-potential critical points are given by the following Lax pair:
\begin{align}
&\left\{
\begin{array}{l}
\ds
\bP (t;\del) = \lambda^{\hat p}\, \bPi^{(\rm real)}(z) \equiv \lambda^{\hat p} \Bigl( \Gamma^{(\rm real)}\, z^{\hat p} 
+ \sum_{n=1}^{\hat p} U_n^{(\rm real)} \,z^{\hat p-n}\Bigr) + O(g_{str}), \cr
\ds
\bQ (t;\del) = \lambda^{\hat q}\, \bXi^{(\rm real)}(z) \equiv \lambda^{\hat q} \Bigl( \Gamma^{(\rm real)}\, z^{\hat q} 
+ \sum_{n=1}^{\hat q} V_n^{(\rm real)} \,z^{\hat q-n} \Bigr) + O(g_{str}), 
\end{array}
\right. \label{DimensionSeparationReal}\\
&\qquad \text{with}\qquad \lambda \equiv t^{\frac{1}{\hat p+\hat q-1}},\qquad 
z= \lambda^{-(\hat p+\hat q)} g_{str}\, t\, \del_t.
\end{align}
with the real-potential gamma matrix $\Gamma^{(\rm real)}$
\begin{align}
\Gamma^{(\rm real)} = 
\begin{pmatrix}
0 & 1 & \cr
 & \ddots   & \ddots \cr
 &    &      0     & 1 \cr
-1&  &             & 0
\end{pmatrix}
\,\simeq\, \omega^{\frac{1}{2}}
\begin{pmatrix}
1 &   \cr
 & \omega   & \cr
 &    &      \ddots     &  \cr
&  &             & \omega^{k-1}
\end{pmatrix}.
\end{align}
Note that all the coefficient matrices $U_n^{(\rm real)}$ and $V_n^{(\rm real)}$ are real functions \cite{CISY1}. 
By taking into account these facts and by using the same procedure as in the previous section, 
one can find the following two types of solutions at the real-potential critical points:
\begin{align}
\underline{\text{cosh solution}}:&\quad
\left\{
\begin{array}{r}
\ds 
\Pi^{(j)}_{\hat p}{}(z) 
= \cosh\bigl(\hat p \tau + 2\pi i \frac{2j-1}{2k}\bigr) 
= T_{\hat p}^{(\frac{2j-1}{2k})}(z), \cr
\ds 
\Xi^{(j)}_{\hat q}{}(z) 
= \cosh\bigl(\hat q \tau + 2\pi i \frac{2j-1}{2k}\bigr) 
= T_{\hat q}^{(\frac{2j-1}{2k})}(z). \label{CoshSolutionReal}
\end{array}
\right. \\
\text{with}\quad &
\det \Bigl[ (2 \zeta ) \,I_k - \bP(z)\Bigr] 
= 2 \lambda^{k\hat p}\Bigl(T_k(\zeta/\lambda^{\hat p}) + T_{k\hat p}(z)\Bigr) =0, \\
\underline{\text{sinh solution}}:&\quad
\left\{
\begin{array}{r}
\ds 
\Pi^{(j)}_{\hat p}{}(z) = \sinh\bigl(\hat p \tau + 2\pi i \frac{2j-1}{2k}\bigr) 
= i \,T_{\hat p}^{(\frac{2j-1}{2k}-\frac{1}{4})}(z), \cr
\ds 
\Xi^{(j)}_{\hat q}{}(z) = \sinh\bigl(\hat q \tau + 2\pi i \frac{2j-1}{2k}\bigr) 
= i \, T_{\hat q}^{(\frac{2j-1}{2k}-\frac{1}{4})}(z). 
\end{array}
\right. \\
\text{with}\quad &
\det \Bigl[ (2 \zeta ) \,I_k - \bP(z)\Bigr] 
= 2 \lambda^{k\hat p}\Bigl((-1)^{\frac{k}{2}}T_k(-i\zeta/\lambda^{\hat p}) + T_{k\hat p}(z)\Bigr) =0. 
\end{align}
Appearance of these solutions also depends on the parity of the number of cuts $k$:
\begin{align}
\text{$k$ is odd}:&\qquad \text{cosh solutions}, \nn\\
\text{$k$ is even}:&\qquad \text{cosh solutions and sinh solutions}. 
\end{align}
In the following, we compare the critical points from two types of the matrix-model potentials with different 
hermiticity:%
\footnote{One way to see this is to draw some figures of phase shifts. 
Other way is to compare the algebraic equations with each other. Here we compare the solutions directly. }
\begin{itemize}
\item When $k$ is odd, 
the cosh solutions of real potentials are equivalent 
to the cosh solution of $\omega^{1/2}$-rotated potentials: 
\begin{align}
\cosh\bigl(\hat p \tau + 2\pi i \frac{2j-1}{2k}\bigr) = - \cosh\bigl(\hat p \tau + 2\pi i \frac{j'-1}{k}\bigr),
\quad (j'-1 = j+\frac{k-1}{2}).
\end{align}
Therefore, these systems are equivalent. 
\item When $k$ is even and $k/2$ is odd, the cosh and sinh solutions 
of real potentials are equivalent 
to the sinh and cosh solutions of $\omega^{1/2}$-rotated potentials, respectively: 
\begin{align}
\cosh\bigl(\hat p \tau + 2\pi i \frac{2j-1}{2k}\bigr) 
&= -i \sinh\bigl(\hat p \tau + 2\pi i \frac{j'-1}{k}\bigr),
\quad (j'-1 = j+\frac{k-2}{4}), \nn\\
\sinh\bigl(\hat p \tau + 2\pi i \frac{2j-1}{2k}\bigr) 
&= i \cosh\bigl(\hat p \tau + 2\pi i \frac{j'-1}{k}\bigr),
\quad (j'-1 = j-\frac{k+2}{4})
\end{align}
Therefore, these systems are also equivalent. 
\item When $k \in 4\mathbb Z$, the cosh solutions and sinh solutions of real potentials are equivalent 
to each other for the same reason as Eq.~\eq{4nmidkSeqC}. 
However the solutions of real potentials are {\em distinct}
from the solutions of $\omega^{1/2}$-rotated potentials. 
\end{itemize}
Note that although these solutions in the real-potential models look similar, 
they are quantitatively {\em distinct} when $k \in 4\mathbb Z$. 
In particular, the $k\in 4\mathbb Z$ cases of $\omega^{1/2}$-rotated-potential 
critical points can be viewed as a mixture of bosonic and superstring system; 
but the $k\in 4\mathbb Z$ cases of real-rotated-potential 
critical points are not. 
Interestingly on the other hand, 
the cosh and sinh solutions of $k\in 4\mathbb Z$ are degenerate
in both $\omega^{1/2}$-rotated-potential and real-potential cases. 
Therefore in this sense, the total number of different solutions are still {\em two} in the $k\in 4\mathbb Z$ cases.

\subsection{Matrix realization of the solutions \label{MatrixRealizationSection}}

We have obtained the solutions by imposing two necessary conditions on the characteristic equations 
of Lax operators. 
In the discussion above, however, no specific form of the matrix-valued Lax pair 
$(\bPi(z),\bXi(z))$ is specified. One may wonder whether some actual expression 
of the Lax operators exists, or how does the expression look like. 

The actual realization of the matrix-valued Lax operators is also non-trivial and important. 
This problem is referred to as {\em matrix realization} of the solutions. 
Purpose of this section is to give explicit matrix realizations of the cosh and sinh solutions. 

\subsubsection{The $\omega^{1/2}$-rotated-potential cases}
In order to identify any matrix realization of the cosh and sinh solutions of 
Eqs.~\eq{CoshSolution} and \eq{SinhSolution}, we need to find out 
a pair of the matrix-valued polynomials $\bPi(z)$ and $\bXi(z)$, 
which commute with each other, $[\bPi(z),\bXi(z)]=0$, and 
whose eigenvalues are given as Eqs.~\eq{CoshSolution} or  \eq{SinhSolution}
of the cosh/sinh solutions. 

While it is possible to find several realizations, 
say for the $(\hat p,\hat q;k)=(1,2;3)$ cases, it turns out that 
a matrix realization of our solution is simply given by 
the following matrices $\bQ^{(n)}_{\pm}(\lambda;z)$:
\begin{align}
\bQ^{(n)}_\pm (\lambda;z)= \lambda^{n}\, \biggl[\Gamma\, T_n(z) 
+& \sum_{s=0}^{\lfloor\frac{n-1}{2}\rfloor}
c_{n-1-2s} \, T_{n-1-2s}(z) \, M_\pm  + \nn\\
+& \sum_{s=0}^{\lfloor\frac{n-2}{2}\rfloor}c_{n-2-2s}\, T_{n-2-2s}(z) \, \bigl(\Gamma\mp \Gamma^{-1}\bigr) \biggr], \label{MatRealComplexCases}
\end{align}
with
\begin{align}
M_{+}\equiv \Delta\,\bigl(\Gamma^l- \Gamma^{l+2}\bigr),\qquad
M_- \equiv  E\Delta\, \bigl(\Gamma^l+\Gamma^{l+2}\bigr). 
\label{MatrixRealizationComplex}
\end{align}
Here $T_n(z)$ is the Chebyshev polynomial of the first kind, and the coefficients $c_n$ 
and the matrix $\Delta$ are defined as 
\begin{align}
c_n\equiv 1- \frac{\delta_{n,0}}{2},\qquad 
\Delta \equiv 
\begin{pmatrix}
 &  &    & 1\cr
 &  & 1 \cr
 & \rotatebox{45}{$\cdots$}&  \cr
1 & & 
\end{pmatrix}. 
\end{align}
Then the claim is that the Lax pair is given by these matrices:
\begin{align}
\text{cosh solutions:}&\qquad 
\bP = \bQ^{(\hat p)}_+,\qquad \bQ = \bQ^{(\hat q)}_+, \nn\\
\text{sinh solutions:}&\qquad 
\bP = \bQ^{(\hat p)}_-,\qquad \bQ = \bQ^{(\hat q)}_-, \label{MatRealClaim}
\end{align}
for every pair of $(\hat p,\hat q)$. 
The integer $l$ is basically a free parameter and is chosen to be an even number when $k$ is even. 
It is because if $k$ is even, there is the residual $\mathbb Z_2$ symmetry \eq{residualZ2Sym} 
and this requires the following condition,
\begin{align}
\{E,M_\pm \}=0, \label{MatrixRealizationSinh}
\end{align}
which results that $l$ should be even. 

For later convenience, 
we collect relevant formulae among the matrices appearing here: 
\begin{align}
\Delta \Gamma^l \Delta = \Gamma^{-l},\qquad \{E,\Delta\}=0,\qquad \{E,\Gamma\}=0,\qquad E^2=I_k. 
\end{align}
In order to check that the matrix-realization \eq{MatRealComplexCases} 
solves the Douglas equation \eq{LeadingDouglasAA},
\begin{align}
[\bQ^{(\hat p)}_\pm,\bQ^{(\hat q)}_\pm] =0,  \label{ComOfMatrixRealization}
\end{align}
for every pair of $(\hat p,\hat q)$, we found that the following condition on $M_\pm$ is sufficient: 
\begin{align}
[\Gamma,M_\pm] \neq 0,\qquad [(\Gamma\pm \Gamma^{-1}),M_\pm ] =0. \label{GammaM}
\end{align}
By repeatedly using these relations \eq{GammaM} 
and the addition formulae of the Chebyshev polynomials 
(or the formulea of hyperbolic cosines), 
one can prove Eq.~\eq{ComOfMatrixRealization}. 
One can verify that the matrices $M_{\pm}$ in Eq.~\eq{MatrixRealizationComplex} 
satisfy this sufficient condition. 

Secondly, one needs to show that 
these matrices in Eq.~\eq{MatRealComplexCases} 
give rise to the algebraic equation for every number of $k$: 
\begin{align}
\det \Bigl[ (2 \zeta ) \,I_k - \bQ^{(\hat p)}_+(z)\Bigr] &
= 2 \lambda^{k\hat p}\Bigl(T_k(\zeta/\lambda^{\hat p}) - T_{k\hat p}(z)\Bigr) =0, \nn\\
\det \Bigl[ (2 \zeta) \,I_k - \bQ^{(\hat p)}_- (z)\Bigr] &
= 2 \lambda^{k\hat p}
\Bigl( (-1)^{\frac{k}{2}}T_k(-i \zeta/\lambda^{\hat p}) - T_{k\hat p}(z)\Bigr) =0. 
\end{align}
Unfortunately, we have not found any general proof of this relation. Instead, 
we have checked this condition by using {\it Mathematica}${}^{\rm TM}$
up to very higher order of $\hat p$ and $k$.%
\footnote{We found several new determinant formulae for the Chebyshev polynomials 
of the first kind, 
\begin{align}
T_k(z) = \frac{1}{2} \det\Bigl[\Gamma z + \frac{1}{2}M_k\Bigr]_{k\times k},\qquad 
T_k(z) -T_k(\zeta) = \frac{1}{2} \det\Bigl[\Gamma z-\zeta I_k + \frac{1}{2}M_k\Bigr]_{k\times k},
\end{align}
which are different from Nash's determinant formula \cite{NashDet}. 
} 

As another non-trivial check, let us consider $k=2$, the two-cut critical points. Since 
the Gamma matrix satisfies $\Gamma^{-1}=\Gamma$, the lax operators can be simplified as%
\footnote{For example, one can use the following formulae: 
$T_n(x) = \frac{1}{2}\Bigl( U_n(x) - U_{n-2}(x)\Bigr)$ and $T_{n+1}(x) = xU_n(x) - U_{n-1}(x)$. }
\begin{align}
\text{cosh solutions:}\qquad  \bQ^{(\hat p)}_+ &= \lambda^{\hat p}\,\Bigl[\Gamma T_{\hat p}(z)\Bigr], \nn\\
\text{sinh solutions:}\qquad  \bQ^{(\hat p)}_- &= \lambda^{\hat p}\, 
\Bigl[\Gamma T_{\hat p}(z) 
+ 2E\Delta
\Bigl(T_{\hat p-1}(z) + T_{\hat p-3}(z)+ \cdots \Bigr) +\nn\\
&\qquad +2 \Gamma
\Bigl(T_{\hat p-2}(z)+ T_{\hat p-4}(z)+\cdots\Bigr) \Bigr]\nn\\
&=\lambda^{\hat p} \, \Bigl[U_{\hat p-1}(z) \bigl(z\Gamma + E\Delta\bigr)\Bigr].
\end{align}
This is identical to the matrix realization given in \cite{CISY1}. 

\subsubsection{The real-potential cases}
The matrix realizations of the real-potential solutions are also obtained as follows:
Despite of the difference in the sinh solutions, actually the matrix realization is 
very similar. What one needs to do is just replace $\Gamma$ by $\Gamma^{(\rm real)}$ 
in the realization of $\omega^{1/2}$-rotated-potential critical points. That is, 
if one introduces the following Lax operators:
\begin{align}
\bQ^{(n)}_{\pm,\, \rm real} = \lambda^{n}\, \biggl[\Gamma^{(\rm real)} \, T_n(z) 
&+ \sum_{s=0}^{\lfloor \frac{n-1}{2}\rfloor}
c_{n-1-2s} \, T_{n-1-2s}(z) \, M_\pm^{(\rm real)}   
+\nn\\
&+ \sum_{s=0}^{\lfloor \frac{n-2}{2}\rfloor}
c_{n-2-2s}\, T_{n-2-2s}(z) \, 
\Bigl(\Gamma^{(\rm real)} \mp \bigl(\Gamma^{(\rm real)}\bigr)^{-1}\Bigr) \biggr],
\end{align}
with
\begin{align}
M_{+}^{(\rm real)}\equiv \Delta\,
\Bigl(\bigr(\Gamma^{(\rm real)}\big)^l- \bigl(\Gamma^{(\rm real)}\bigr)^{l+2}\Bigr), \quad 
M_-^{(\rm real)} \equiv E\Delta\,\Bigl(\bigl(\Gamma^{(\rm real)}\bigr)^l+\bigl(\Gamma^{(\rm real)}\bigr)^{l\pm 2}\Bigr),
\end{align}
the matrix realization of each solution is given as 
\begin{align}
\text{cosh solutions:}&\qquad 
\bP = \bQ^{(\hat p)}_{+,\, \rm real},
\qquad 
\bQ = \bQ^{(\hat q)}_{+,\, \rm real}, \nn\\
\text{sinh solutions:}&\qquad 
\bP = \bQ^{(\hat p)}_{-,\, \rm real},
\qquad 
\bQ = \bQ^{(\hat q)}_{-,\,\rm real}.
\end{align}
For further illustrations, we list some of the examples for $M_{\pm}$ in Appendix \ref{ExampleMrealization}. 

\section{Algebraic geometry of fractional superstrings \label{AlgebraicStructuresSection}}

In this section, we study the algebraic curves of the cosh and sinh solutions. The central object is 
the algebraic equations $F(\zeta,Q)=0$ of macroscopic loop amplitudes $Q(\zeta)$. As a definition 
of algebraic equations, we adopt the definition given in the free-fermion formulation \cite{fim}, especially 
its multi-component extension \cite{fi1}, 
where the form of the equation is required from the $W_{1+\infty}$-constraints, i.e.~the Loop equations 
of the matrix models.%
\footnote{It should be noted that, before \cite{fim,fi1}, the form of the algebraic equation itself 
was found in \cite{EynardAlg} and used in \cite{KazakovKostov} to show the relationship 
with the algebraic equation given in Liouville side \cite{SeSh}. }
The definition is given as follows:

Macroscopic loop amplitudes $Q(\zeta)$ and the eigenvalues of the Lax operators $(\bP,\bQ)$ 
are now related as
\begin{align}
Q(\zeta) 
\qquad \Leftrightarrow\qquad 
\left\{
\begin{array}{l}
\ds
\zeta = \lambda^{\hat p}\,\Pi_{\hat p}^{(j)}(z),\cr 
\ds
Q = \lambda^{\hat q}\, \Xi_{\hat q}^{(j)}(z).
\end{array}
\right. \qquad (j=1,2,\cdots,k). \label{SolutionsInSecAlgEq}
\end{align}
The index $j$ is the label of the eigenvalues $j=1,2,\cdots,k$. 
Since there are many $Q^{(j)}(\zeta)$ corresponding to the single $Q(\zeta)$, 
the algebraic equation $F(\zeta,Q)=0$ is defined so that it includes 
{\em all the eigenvalues $Q^{(j)}(\zeta)$ as its solutions} \cite{fi1}.

The macroscopic loop amplitude $Q^{(j)}(\zeta)$ is now written as 
\begin{align}
\text{cosh solutions:} & \quad
Q^{(j)}(\zeta) 
= \lambda^{\hat q}\,T_{\hat q/\hat p}^{(- \frac{(\hat q-\hat p)}{\hat p} \nu_j)}(\zeta/\lambda^{\hat p}), \nn\\
\text{sinh solutions:} & \quad
Q^{(j)}(\zeta) 
= i \,\lambda^{\hat q}\,T_{\hat q/\hat p}^{
(- \frac{(\hat q-\hat p)}{\hat p}[\nu_j- \frac{1}{4}])
}(-i \zeta/\lambda^{\hat p}), \label{SolQofJ}
\end{align}
where the phase shifts $\nu_j$ are given as
\begin{align}
\nu_j = 
\left\{
\begin{array}{ll}
\ds
\frac{(j-1)}{k} & : \text{$\omega^{1/2}$-rotated potentials} \cr
\ds
\frac{(2j-1)}{2k} & : \text{real potentials}.
\end{array}
\right. 
\end{align}
with $j=1,2,\cdots,k$. 
By introducing the $\hat p$ branches of the loop amplitudes $Q_a^{(j)}(\zeta)$ \cite{fkn},%
\footnote{By construction, $Q_0^{(j)}(\zeta)=Q^{(j)}(\zeta)$ 
and $Q_{a+\hat p}^{(j)}(\zeta) = Q_a^{(j)}(\zeta)$. }
\begin{align}
Q_a^{(j)}(\zeta) \equiv 
\left\{
\begin{array}{ll}
\lambda^{\hat q}\,T_{\hat q/\hat p}^{
(- \frac{(\hat q-\hat p)}{\hat p}\nu_j+\frac{a}{\hat p})
}(\zeta/\lambda^{\hat p}) &
\text{(cosh solutions)}, \cr
i \,\lambda^{\hat q}\,T_{\hat q/\hat p}^{
(- \frac{(\hat q-\hat p)}{\hat p}[\nu_j- \frac{1}{4}]+\frac{a}{\hat p})
}(-i \zeta/\lambda^{\hat p}) & 
\text{(sinh solutions)}, \label{QofZetaPlusA}
\end{array}
\right.
\end{align}
the algebraic equations are defined as 
\begin{align}
F(\zeta,Q) 
\equiv \prod_{j=1}^{k}
\Bigl[\prod_{a=0}^{\hat p-1} 
\Bigl(Q - Q_a^{(j)}(\zeta)
\Bigr)
\Bigr] 
=0.
\end{align}
In the case of $k=1$, this goes back to the expression \cite{EynardAlg,KazakovKostov} (the one-cut two-matrix models) and \cite{fim} (the $W_{1+\infty}$ approach). Generally this is the definition given in 
\cite{fi1} (the multi-component cases). 

Organization of this section is following: 
Algebraic equations are in section \ref{AlgebraicEquationsSection}, 
branch points are in section \ref{BranchCurvesSection}, 
the multi-cut geometry is in section \ref{MultiCutGeometryStokesSection} 
and singular points are in section \ref{AlgSingularPointsSection}. 

\subsection{Algebraic equations and 
the $\mathbb Z_k \times \mathbb Z_k$ symmetry \label{AlgebraicEquationsSection}}

Let us obtain the algebraic equations of our solutions. 
First consider the multiplication among different branches with a fixed eigenvalue $j$, 
which results in 
\begin{align}
&\prod_{a=0}^{\hat p-1} 
\Bigl(
Q - \lambda^{\hat q}\,T_{\hat q/\hat p}^{
(- \frac{(\hat q-\hat p)}{\hat p}\nu_j+\frac{a}{\hat p})
}(\zeta/\lambda^{\hat p})
\Bigr) = 
\frac{\lambda^{\hat p/\hat q}}{2^{\hat p-1  }} 
\times \Bigl(T_{\hat p}(Q/\lambda^{\hat q})
-T_{\hat q}^{(-(\hat q-\hat p)\nu_j)}(\zeta/\lambda^{\hat p})\Bigr), \nn\\
&\prod_{a=0}^{\hat p-1} 
\Bigl(
Q - i\lambda^{\hat q}\,T_{\hat q/\hat p}^{
(- \frac{(\hat q-\hat p)}{\hat p}[\nu_j- \frac{1}{4}]+\frac{a}{\hat p})
}(-i \zeta/\lambda^{\hat p})
\Bigr) = \nn\\
&\qquad =\frac{\lambda^{\hat p/\hat q}\,  i^{\hat p} }{2^{\hat p-1  }}
\times \Bigl(T_{\hat p}(-i Q/\lambda^{\hat q})
-T_{\hat q}^{(-(\hat q-\hat p)[\nu_j-\frac{1}{4}])}(-i\zeta/\lambda^{\hat p})\Bigr). 
\end{align}
Introducing the following integers $d_{\hat q-\hat p}, \,\hat k$ and $\eta$ \cite{irie2}, 
\begin{align}
d_{\hat q-\hat p} \equiv \text{g.c.d}\bigl\{\hat q-\hat p, k\bigr\},\qquad 
k = \hat k \times d_{\hat q-\hat p}, \qquad \hat q-\hat p = \eta \times d_{\hat q-\hat p},
\end{align}
one can rewrite $(\hat q-\hat p)\nu_j$ as 
\begin{align}
\text{real potentials:}\qquad 
(\hat q-\hat p)\nu_j &= (\hat q-\hat p) \frac{j-1}{k} + \Bigl\{ \frac{\hat q-\hat p}{2k}\Bigr\} 
= \frac{\eta(j-1)}{\hat k} + \Bigl\{ \frac{\eta}{2\hat k}\Bigr\} \nn\\
&= \frac{r_j}{\hat k}+ l_j + \Bigl\{ \frac{\eta}{2\hat k}\Bigr\}\qquad \bigl(\eta (j-1) = \hat k l_j + r_j\bigr),
\end{align}
and similarly, 
\begin{align}
\text{$\omega^{1/2}$-rotated potentials:}\qquad 
(\hat q-\hat p)\nu_j = 
\frac{r_j}{\hat k}+ l_j \qquad \bigl(\eta (j-1) = \hat k l_j + r_j\bigr). 
\end{align}
Here $\{...\}$ is the part which appears only in the real-potential models. 
Since $\hat k$ and $\eta$ are coprime integers, the integer $r_j$ takes 
each value $(r_j= 1,2,\cdots,\hat k)$ exactly $d_{\hat q-\hat p}$ times. 
Considering $T_n^{(\nu)}(z)=T_n^{(\nu+\mathbb Z)}(z)$, 
one concludes that the algebraic equations of the cosh and sinh solutions 
at the $(\hat p,\hat q;k)$ critical points are expressed as 
\begin{align}
&\underline{\text{cosh solutions}}: \nn\\
&\quad F(\zeta,Q)  = \Bigl(\frac{\lambda^{p/\hat q}}{2^{p-1  }} \Bigr)^{d_{\hat q-\hat p}}
\times \Bigl(T_p(Q/\lambda^{\hat q})
-(-1)^{\{ \eta \}}\,T_q(\zeta/\lambda^{\hat p})\Bigr)^{d_{\hat q-\hat p}} = 0, \nn\\
&\underline{\text{sinh solutions}}: \nn\\
&\quad F(\zeta,Q)  =  \Bigl(\frac{ i^p\lambda^{p/\hat q}}{2^{p-1  }} \Bigr)^{d_{\hat q-\hat p}} 
\times \Bigl(  T_p(-i Q/\lambda^{\hat q}) 
- (-1)^{\{\eta\}+\frac{q-p}{2}} T_q(-i \zeta/\lambda^{\hat p}) \Bigr)^{d_{\hat q-\hat p}} = 0. \label{OurCurveMat}
\end{align}
Here again $\{...\}$ is the part which appears only in the real-potential models. 
We have introduced the $(p,q)$ labeling of 
minimal fractional superconformal field theory \cite{irie2}, 
\begin{align}
(p,q) = (\hat k \hat p,\hat k \hat q). 
\end{align}
Note that this factorization property, $(...)^{d_{\hat q-\hat p}}$, can be understood as 
a consequence of the accidental {\em $\mathbb Z_k \times \mathbb Z_k$ symmetry} of this background. 
That is, the $k$-cut two-matrix models include $d_{\hat q-\hat p}$ copies of $(\hat p,\hat q;k)$ 
minimal fractional superstring theory and the single minimal fractional superstring theory can be singled out after 
imposing $\mathbb Z_k\times \mathbb Z_k$ gauging of the system \cite{irie2}.%
\footnote{This concept was first introduced in \cite{fi1} 
in order to obtain $(\hat p,\hat q)$ odd models of minimal superstrings from the two-cut two-matrix models. } 
In this sense, the cosh and sinh solutions are consistent 
with the $\mathbb Z_k \times \mathbb Z_k$ gauging requirement 
which has been anticipated from the Liouville side. Since these copies become {\em identical}
after the $\mathbb Z_k \times \mathbb Z_k$ gauging, 
the algebraic equations of the Liouville side, $F_L(\zeta,Q) =0 $, are given as
\begin{align}
\underline{\text{cosh solutions}}: &\quad F_L(\zeta,Q)  
= T_p(Q/\lambda^{\hat q}) -(-1)^{\{ \eta \}}\,T_q(\zeta/\lambda^{\hat p}) = 0, \nn\\
\underline{\text{sinh solutions}}: &\quad F_L(\zeta,Q)  
=  T_p(-i Q/\lambda^{\hat q}) 
- (-1)^{\{\eta\}+\frac{q-p}{2}} T_q(-i \zeta/\lambda^{\hat p}) = 0. \label{OurCurves}
\end{align}

\subsection{Branch points of the curves \label{BranchCurvesSection}}

Let first consider branch points of the curves. Branch points here are defined as 
$\dfrac{\del Q(\zeta)}{\del \zeta} = \infty$. Our current purpose is to see where the branch points 
exist on the branch \eq{SolQofJ}. These branches (or solutions) \eq{SolQofJ} 
of the algebraic equation are particularly important because they correspond to macroscopic loop amplitudes
\eq{SolutionsInSecAlgEq}. Therefore, we here refer to these branches as {\em physical branches}. 

For sake of simplicity, we put $\lambda=1$ and consider the following loop amplitudes:
\begin{align}
\zeta = \cosh(\hat p \tau + 2\pi i \nu),\qquad Q^{(j)} = \cosh(\hat q \tau + 2\pi i \nu). 
\end{align}
The branch point of this expression is obtained from 
\begin{align}
\infty=\dfrac{\del Q^{(j)}(\zeta)}{\del \zeta} 
=\frac{\hat q}{\hat p}\, \frac{\sinh(\hat q \tau + 2\pi i \nu )}{\sinh(\hat p \tau + 2\pi i \nu)} . 
\end{align}
Since the function $\sinh \tau$ is finite on $\tau \in \mathbb C$, 
the branch points exist only if 
\begin{align}
\sinh(\hat p \tau + 2\pi i \nu)= 0 &\quad \Leftrightarrow \quad \tau 
= \pi i \Bigl[-\frac{m+2\nu}{\hat p}\Bigr]\quad \Leftrightarrow\quad \zeta = (-1)^m, \nn\\
\sinh(\hat q \tau + 2\pi i \nu)\neq 0 &\quad \Leftrightarrow \quad \tau 
\neq \pi i \Bigl[\frac{n-2\nu}{\hat q}\Bigr]\quad \Leftrightarrow\quad Q^{(j)}\neq (-1)^n,
\end{align}
with $m,n\in \mathbb Z$. 
That is, there are generally {\em two} branch points $\zeta=\pm 1$ on this branch $Q^{(j)}(\zeta)$. 

\begin{figure}[htbp]
 \begin{center}
  \includegraphics[scale=0.75]{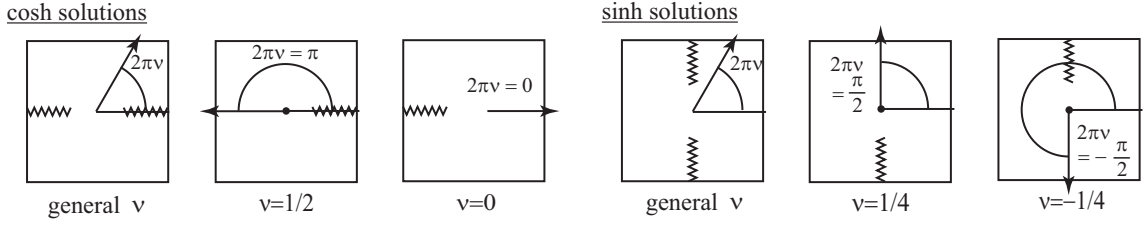}
 \end{center}
 \caption{\footnotesize 
The positions of cuts in the physical branches. 
In the cosh solutions, the branches of $\nu=0,1/2$ only include 
one cut and exactly the same branch as bosonic string. The cases of $\nu=\pm 1/4$ are the same as the two-cut 
phase of type 0 superstrings; In sinh solutions, the branches of $\nu=\pm 1/4$ only include 
one cut and exactly the same branch as bosonic string. The cases of $\nu=0,1/2$ are the same as the two-cut 
phase of type 0 superstrings. In all the branches, there are maximally two branch points on each sheet. } 
 \label{physbranch}
\end{figure}

It is also convenient to consider when one of the two points $\zeta=\pm1$ becomes 
a normal point: it happens if they satisfy
\begin{align}
\tau = \pi i \Bigl[-\frac{m+2\nu}{\hat p}\Bigr] = \pi i \Bigl[\frac{n-2\nu}{\hat q}\Bigr]  
\quad &\Leftrightarrow\quad  
\hat q m+\hat p n = 2 \nu (\hat p-\hat q).
\end{align}
Therefore, the condition that one of the branch points disappears is given as 
\begin{align}
\text{cosh solution:}&\qquad 
(\hat q-\hat p) \Bigl[\frac{j-1}{k}\Bigr]
= \eta \frac{j-1}{\hat k} \in \frac{1}{2} \mathbb Z, \nn\\
\text{sinh solution:}&\qquad 
(\hat q-\hat p) \Bigl[\frac{j-1}{k} -\frac{1}{4}\Bigr]
= \eta \frac{j-1}{\hat k} - \frac{\hat q-\hat p}{4} \in \frac{1}{2} \mathbb Z. 
\end{align}
This indicates that 
the branch points only appear when $\zeta= \pm i, Q\neq \pm i$. 
The positions of cuts on these branches for the unitary cases $\hat q-\hat p=1$ are shown in Fig.\ref{physbranch}. 

Note that the curves essentially depend on the effective phase shift $\hat \nu_j = (j-1)/\hat k$. 
These curves typically appear in the $\hat k$-cut matrix models or $\hat k$-fractional superstring theory. 
Here we refer to these curves as {\em $F_{\hat \nu}$-string} curves. For example, $F_{0}$- and $F_{\pm 1/2}$-
string curves are equivalent to the curves of bosonic minimal string theory, and $F_{\pm 1/4}$-string curves
are equivalent to minimal type 0 superstring theory. 
Also note that one can also consider dual branch points $\dfrac{\del \zeta(Q)}{\del Q}=\infty$, 
which result in the opposite condition, $\zeta \neq (-1)^m,\ Q^{(j)} = (-1)^n$. 

\subsection{The multi-cut geometry and Stokes phenomenon \label{MultiCutGeometryStokesSection}}

In the previous subsection, we have obtained branch points of 
the physical branches of macroscopic loop amplitudes, $Q^{(j)}(\zeta)$, and 
it turns out that they include at most {\em two} cuts on each branch. 
Since we have started our discussion from the $k$-cut matrix models, 
there should be $k$ cuts in the macroscopic loop amplitudes $Q(\zeta)$. That is, 
{\em where is the multi-cut geometry?} 
A simple answer to this question is the follwing: 
The $k$ cuts exist on {\em physical sheet} of the $k$-cut matrix models. 

So far, we have obtained eigenvalues 
$\bigl(\lambda^{\hat p}\Pi_{\hat p}^{(j)}(z), \lambda^{\hat q}\Xi_{\hat q}^{(j)}(z)\bigr)$ 
of the Lax operators $(\bP,\bQ)$ in the weak coupling limit, 
\begin{align}
\zeta\, \Psi^{(j)}(t;\zeta) 
&= \bP(t;\del)\, \Psi^{(j)}(t;\zeta) 
= \lambda^{\hat p}\, \Pi^{(j)}_{\hat p}(z) \, \Psi^{(j)}(t;\zeta) + O(g_{\rm str}), \nn\\
g_{\rm str}\, \frac{\del}{\del \zeta}\, \Psi^{(j)}(t;\zeta) 
&=\bQ(t;\del)\, \Psi^{(j)}(t;\zeta)
= \lambda^{\hat q}\, \Xi^{(j)}_{\hat q}(z) \, \Psi^{(j)}(t;\zeta) + O(g_{\rm str}), \label{BakAkhStokes}
\end{align}
and calculated the corresponding algebraic equations \eq{OurCurveMat}. 
As one may notice, these algebraic curves are highly reducible. 
For example, the algebraic equation of cosh solutions,%
\footnote{The sinh case is almost the same except for replacing $(\zeta,Q) \to (-i\zeta,-i Q)$. }
can be factorized in the following way:
\begin{align}
F(\zeta,Q) = \prod_{j=1}^{\lfloor \frac{k}{2}\rfloor} F^{(\nu_j)}(\zeta,Q)=0, \label{FactorizationCurve1}
\end{align}
where $\nu_j$ is the phase shift of the solution, 
and $F^{(\nu_j)}(\zeta,Q)=0$ is the irreducible algebraic equation of an eigenvalue $Q^{(j)}(\zeta)$ of 
the Lax operators:
\begin{align}
F^{(0)}(\zeta,Q) 
&=F^{(1/2)}(-\zeta,-Q)
= T_{\hat p}(Q/\lambda^{\hat q})-T_{\hat q}(\zeta/\lambda^{\hat p}) = 0, \nn\\
F^{(\nu)}(\zeta,Q) 
&= \Bigl[ T_{\hat p}(Q/\lambda^{\hat q})-T_{\hat q}^{((\hat q-\hat p)\nu)}(\zeta/\lambda^{\hat p})\Bigr]
\Bigl[ T_{\hat p}(Q/\lambda^{\hat q})-T_{\hat q}^{(-(\hat q-\hat p)\nu)}(\zeta/\lambda^{\hat p})\Bigr] =0. 
\label{FactorizationCurve2}
\end{align}
This is the origin of the fact that 
there are maximally {\em two} cuts 
on the physical branch \eq{SolQofJ} and the $k$ necessary cuts are missing. 

Here we have used two different words {\em physical branch} and {\em physical sheet}. 
Although the word, physical sheet, is enough to explain everything in the usual one-cut cases,
the multi-cut matrix models require to distinguish them. {\em Physical sheet}, $\mathbb C_{\rm phys}$, 
is a complex plane $\mathbb C$ on which the resolvent $\del_x \Omega(x)$ 
(or macroscopic loop amplitude $Q(\zeta)$) is defined and each weak coupling infinity 
(say, $x \to e^{2\pi i \theta_j} \times \infty;\, j=1,2,\cdots,k$) gives correct asymptotic 
behavior:
\begin{align}
\del_x \Omega(x) = \vev{\frac{1}{N}\tr \frac{1}{x-X}}\to \frac{1}{x},\qquad 
x\, (\in \mathbb C_{\rm phys}) \to e^{2\pi i \theta_j} \times \infty. \label{PhysicalSheet}
\end{align}
Therefore, in the $k$-cut matrix models, $Q(\zeta)$ on the physical sheet is expected 
to have $k$ semi infinite cuts which separate the $k$ weak coupling regimes. 

On the other hand, if one solves the algebraic equation \eq{OurCurveMat}, 
generally $k$ solutions (i.e.~branches) satisfy the condition \eq{PhysicalSheet} with some particular value of $j$. 
Due to the definition \eq{BakAkhStokes}, 
they are naturally identified with $Q^{(j)}(\zeta)$ $(j=1,2,\cdots,k)$. 
Therefore, these solutions or branches are called {\em physical branch}. 
Physical sheet and physical branch are equivalent concepts in the one-cut cases. 
Note that only the cuts on physical sheets have the physical meaning as condensation 
of the matrix-model eigenvalues $\{x_i\}_{i=1}^N$ and the position of the cuts can have the geometrical meaning.  

Mathematically, position of cuts is meaningless and one can freely move it continuously. 
Because of this fact, when some of the cuts on physical sheet have no branch point, 
the physical sheet is factorized into several pieces of curves, and 
these cuts on physical sheet disappear from the viewpoint of branch points in the algebraic equation. 
This usually happens when branch points are joined and several cuts are connected. 
Examples are found in the one-cut phase of the two-cut matrix models \cite{UniCom} 
and also ``one-cut phase'' counterpart of the $k$-cut two-matrix models of the $\mathbb Z_k$ symmetric critical points \cite{CISY1}. 
In these cases, although the cuts are missing from the algebraic equations, 
one has to take into account the position of the cuts because they have the definite physical meaning. 

\begin{figure}[htbp]
 \begin{center}
  \includegraphics[scale=0.7]{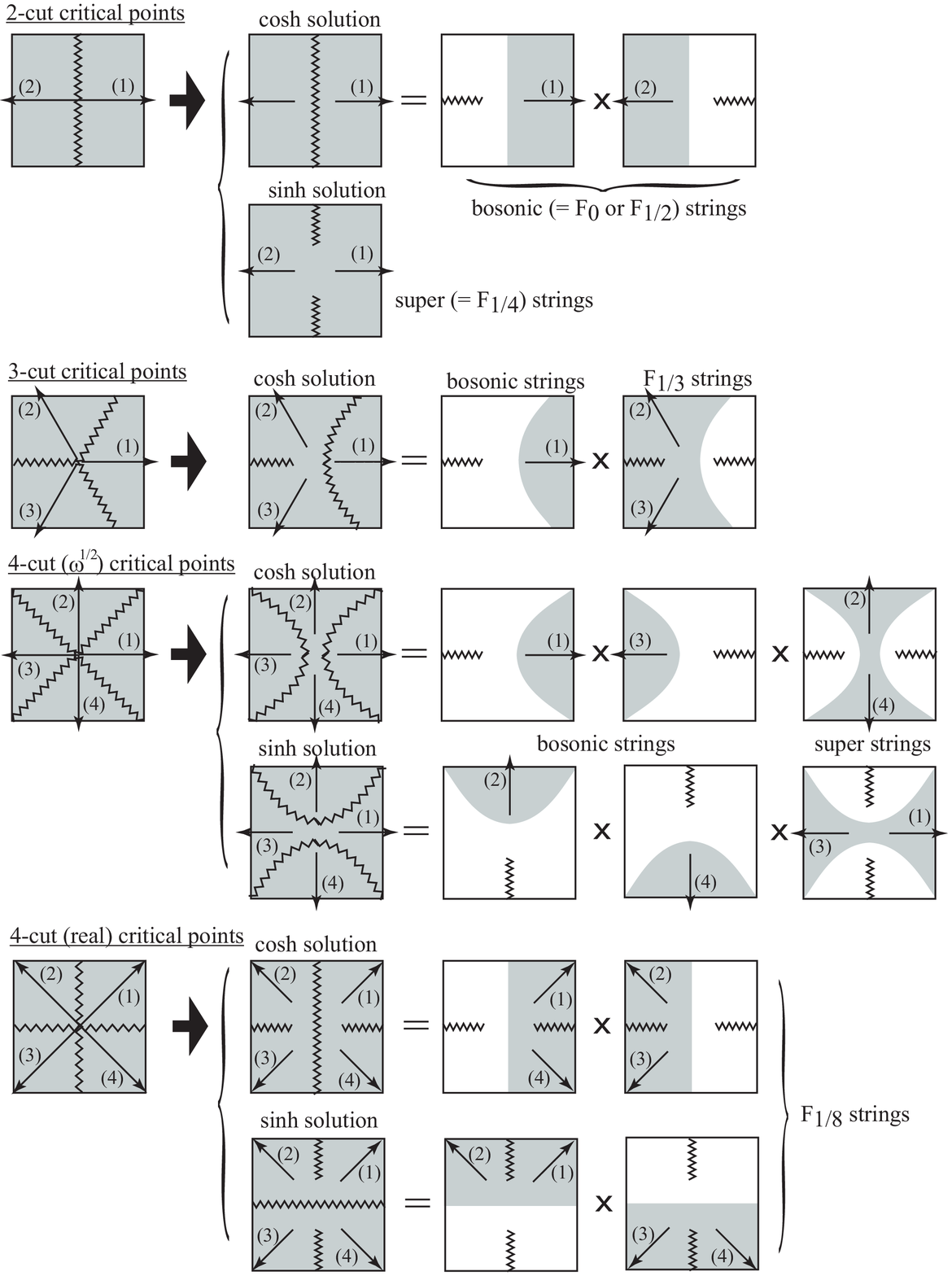}
 \end{center}
 \caption{\footnotesize 
The multi-cut curves in 2,3,4-cut cases. Shaded parts are physical sheets. 
Arrows indicate the weak string coupling regimes of $\zeta$ ($\arg (\zeta)= 2\pi \nu_j;\, j=1,2,\cdots$). The physical branch of each irreducible curve is shown,
which corresponds to some fractional superstring theory of $\nu_j=(j-1)/k$ (denoted as $F_{\nu_j}$-strings). 
Note that this is the geometry of unitary cases $\hat q-\hat p=1 \Rightarrow \hat k=k$. }
 \label{fig1}
\end{figure}

\begin{figure}[htbp]
 \begin{center}
  \includegraphics[scale=0.7]{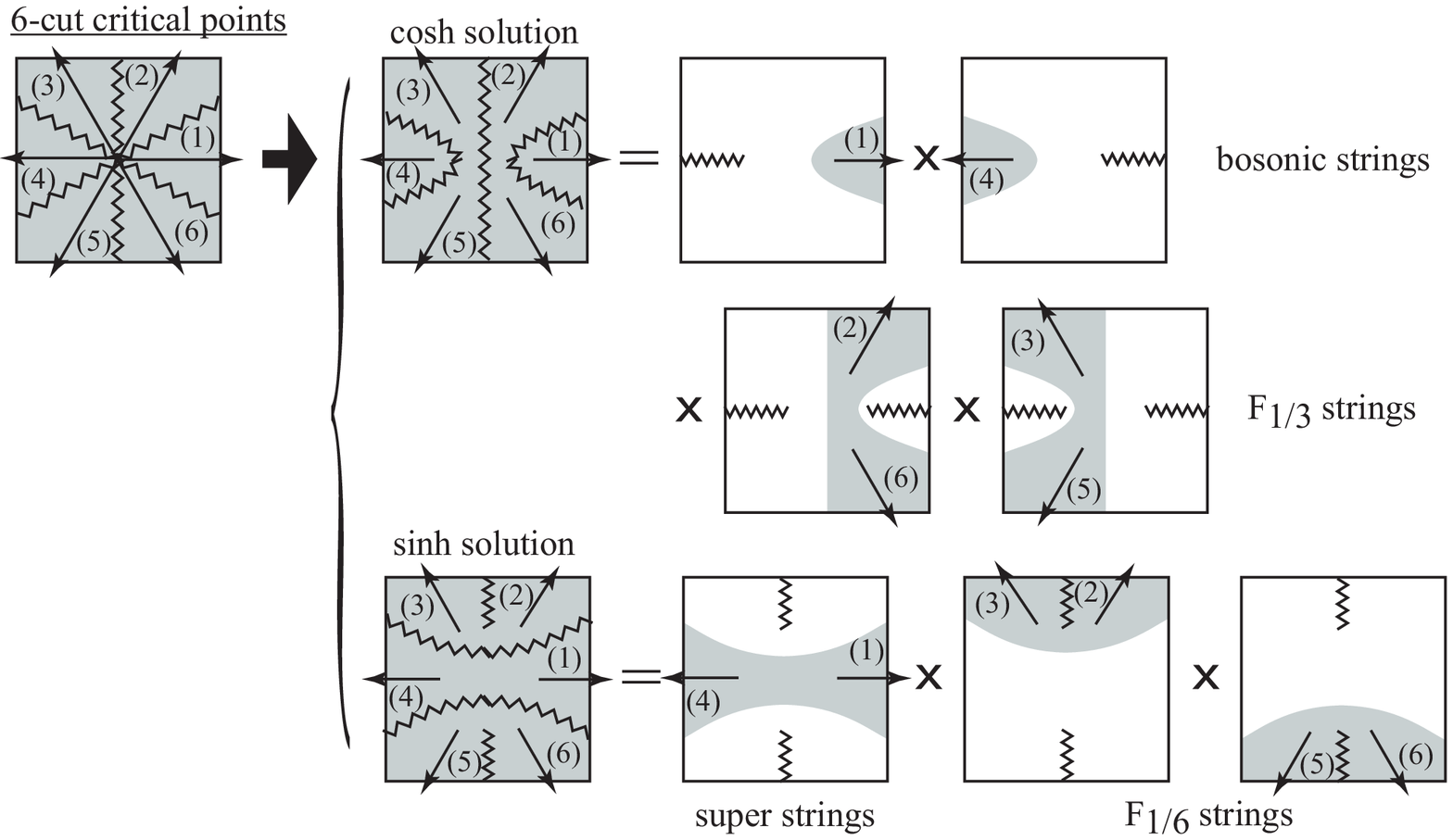}
 \end{center}
 \caption{\footnotesize 
The multi-cut curves in 6-cut cases. Shaded parts are physical sheets. 
Arrows indicate the weak string coupling regimes of $\zeta$ ($\arg (\zeta)= 2\pi \nu_j;\, j=1,2,\cdots$). 
The physical branch of each irreducible curve is shown,
which corresponds to some fractional superstring theory of $\nu_j=(j-1)/k$ (denoted as $F_{\nu_j}$-strings).  
Note that this is the geometry of unitary cases $\hat q-\hat p=1 \Rightarrow \hat k=k$. }
 \label{fig2}
\end{figure}

In our fractional-superstring solutions, the curves are factorized and many cuts are obviously missing. 
This means that we need to add the cuts and glue the physical branches to make up the physical sheet. 
Identification of these cuts generally requires non-perturbative analysis, but at least in the unitary cases
$\hat q-\hat p =1$,%
\footnote{The sinh solutions become subtle when $d_{\hat q-\hat p} \in 2\mathbb Z$, 
which includes $(\hat p,\hat q)$ odd models of type 0 superstring theory. }
the following connection rule of the macroscopic loop amplitude $Q(\zeta)$ on the physical sheet 
(with the $k$ eigenvalues $Q^{(j)}(\zeta)$ of Eq.~\eq{BakAkhStokes}) 
is consistent with the multi-cut geometry: 
\begin{align}
\vev{\frac{1}{N}\tr \frac{1}{x-X}} \quad 
\leftrightarrow \quad 
Q(\zeta) = Q^{(j)}(\zeta) \quad \text{in \,
$2\pi \nu_j -\frac{\pi }{k} < \arg (\zeta) <  2\pi \nu_j +\frac{\pi }{k}$}. 
\label{ConnectionRelation}
\end{align}
That is, $Q^{(j)}(\zeta)$ is the solution of the weak-coupling regime
$\zeta \to \infty \times e^{2\pi i \nu_j}$. 
Therefore, we propose it as a conjecture which can be checked by some non-perturbative analysis. 
We show several examples of the positions of the cuts in Fig.~\ref{fig1} and Fig.~\ref{fig2}. 

Note that, in the $\omega^{1/2}$-rotated-potential cases, 
the coordinate $\zeta$ is rotated by $\omega^{1/2}$ in the complex plane (see Eq.~\eq{RelXandZeta}). 
Therefore, for example the cuts on the physical sheet in the two-cut matrix models 
should only appear along the pure-imaginary axis, although the cosh solutions only 
have cuts along the real axis. 
Our solution to this issue is that they are hided by another physical branch 
and do not appear on the physical sheet. 
In this way, the connection relation \eq{ConnectionRelation} 
of the macroscopic loop operator (or resolvent) provides consistent 
relationship between the physical branches 
and the multi-cut geometries. 
Although we cannot deny the possibility that there is another meaningful connection rules, but 
we leave it as futur works. 
Also note that we put each irreducible curve some particular names, 
as $F_{\nu}$-strings. 

As one sees in Eq.~\eq{ConnectionRelation}, 
these macroscopic loop amplitudes have many non-trivial jumps on 
the physical sheet. This is understood as {\em Stokes phenomenon}%
\footnote{The Stokes phenomenon usually means that a single analytic function $\Psi(\zeta)$ 
can have different asymptotic forms with respect to different angular regimes of $\zeta$ 
if the expansion is not a convergent series. Each angular sector is now called a Stokes sector. 
The claim is that Stokes sectors of the Baker-Akhiezer functions are regimes between cuts of the matrix models.} 
of the Baker-Akhiezer functions \cite{MMSS,SeSh2}: 
Asymptotic behavior of the Baker-Akhiezer functions 
($\zeta \to \infty$) can be changed by an analytic continuiation of $\zeta$ which crossese the cuts 
on the physical sheet. 
Therefore, the relation \eq{ConnectionRelation} indicates how to connect there Stokes sectors 
in the non-perturbative Baker-Akhiezer function. 

From the definition \eq{BakAkhStokes} and the relation \eq{ConnectionRelation}, 
the leading behaviors of the recursive equations can be solved as
\begin{align}
\Psi^{(j)}(t;\zeta) = \exp\Bigl[ \frac{1}{g_{\rm str}}\, \int^{\zeta} \, d\zeta' \, Q^{(j)}(\zeta') \Bigr]\,
\chi^{(j)}(t;\zeta),  \label{AsymPsiJei}
\end{align}
in $\ds 2\pi \nu_j -\frac{\pi }{k} < \arg (\zeta) <  2\pi \nu_j +\frac{\pi }{k}$. 
Here $\chi^{(j)}(t;\zeta)$ is a rank $k$ vector-valued function of sub leading corrections, 
and the expansion is carried out in some weak coupling regimes of the cosmological constant $\zeta$: 
$\zeta \to e^{2\pi i \nu_l} \times \infty$. 
Since the orthonormal polynomials are now related to the macroscopic loop amplitudes, 
\begin{align}
\Psi(t;\zeta)\quad  \leftrightarrow \quad  \alpha_n(x),
\qquad \Bigl(\, \text{i.e. } Q(\zeta) \quad \leftrightarrow \quad \del_x \Omega(x) = \vev{\frac{1}{N}\tr \frac{1}{x-X}}\Bigr), 
\end{align}
these Baker-Akhiezer functions are connected to each other by an analytic continuation:%
\footnote{Note that, in the case of two-cut $(\hat p,\hat q)=(1,2)$, 
this relation has been shown to be true in \cite{SeSh2} (i.e.~Eq.~(4.16))
by using the non-perturbative analysis of the Baker-Akhiezer functions \cite{BakAkhNonPerTwoCut}. } 
\begin{align}
\Psi(t;\zeta) = \Psi^{(j)}(t;\zeta) \qquad 
\text{in \,
$2\pi \nu_j -\frac{\pi }{k} < \arg (\zeta) <  2\pi \nu_j +\frac{\pi }{k}$}. 
\label{StokesBakerAnaly}
\end{align}
Another implication on the free fermion system is also noted in Appendix \ref{ZkChargeFreeFermion}. 

\subsection{Singular points and intersection of sheets \label{AlgSingularPointsSection}}

In this subsection, we study singular points of the cosh and sinh solutions \eq{OurCurves}. 
The definition of singular points is
\begin{align}
F(\zeta,Q)= \frac{\del F(\zeta,Q)}{\del Q}
= \frac{\del F(\zeta,Q)}{\del \zeta} = 0. \label{SingularBranches}
\end{align}
One of the easy ways to obtain these points is 
to consider coincident points among the different branch expressions, 
$Q_a^{(j)}(\zeta)$, that is, the points $(\zeta,Q)$ which satisfy 
\begin{align}
Q = Q_a^{(j)}(\zeta)= Q_b^{(j')}(\zeta), \label{SingularBranchesCond1}
\end{align}
and Eq.~\eq{SingularBranches}. This identification of singular points 
has been used in \cite{KazakovKostov}\cite{fis,fim,fi1}. 
Note that the pair $(j,a)$ cannot be the same as $(j',b)$. 
Or equivalently, we need to solve for the following problem of finding the pair of $(\tau,\tau')$:
\begin{align}
\zeta/\lambda^{\hat p} 
= \cosh \bigl(\hat p \tau + 2\pi i \nu\bigr) 
= \cosh \bigl(\hat p \tau' + 2\pi i \nu'\bigr),\nn\\
Q/\lambda^{\hat q} 
= \cosh \bigl(\hat q \tau + 2\pi i \nu\bigr) 
= \cosh \bigl(\hat q \tau' + 2\pi i \nu'\bigr), \label{SingularBranchesCond2}
\end{align}
where
\begin{align}
(\nu,\nu')=(\nu_j,\nu_{j'}),\qquad
\nu_j = 
\left\{
\begin{array}{ll}
\ds
\frac{j-1}{k} & \text{: $\omega^{1/2}$-rotated potentials}, \cr
\ds
\frac{2j-1}{2k} & \text{: real potentials},
\end{array}
\right.
\end{align}
with $j=1,2,\cdots,k$. 
In the case of the sinh solutions, the parameters $\nu$ and $\nu'$ are shifted by $1/4$:%
\footnote{Or one can say they are a $\pi/2$ shift:
$2\pi \nu \to 2\pi \nu - \frac{\pi}{2}$, $2\pi \nu' \to 2\pi \nu' - \frac{\pi}{2}$.}
\begin{align}
\nu \to \nu - \frac{1}{4},\qquad \nu' \to \nu' - \frac{1}{4}, 
\end{align}
and $(\zeta,Q)$ is replaced by $(-i \zeta,-i Q)$. 
Actually the discussion below does not depend on whether the solutions are cosh or sinh solutions 
and whether they come from $\omega^{1/2}$-rotated potentials or real potentials. 
Therefore, we assume that $\nu$ is a general rational number with a periodicity identification 
$\nu \sim \nu + \mathbb Z$, that is,
\begin{align}
\nu \in \mathbb Q,\qquad 0\leq \nu < 1. 
\end{align}
If one takes $k\to \infty$ limit, the parameter $e^{2\pi i \nu}$ will take the value in $U(1)$. 

The solutions to Eq.~\eq{SingularBranchesCond2} are obtained as 
the following pair of $(\tau,\tau')$, which are parameterized by a pair of integers, $(m,n)$: 
\begin{align}
\tau_{m,n} &= \pi i \Bigl[- \frac{1}{\hat p}\Bigl(m-(\nu'-\nu)\Bigr) 
+ \frac{1}{\hat q}\Bigl(n-(\nu'+\nu)\Bigr)\Bigr],\nn\\
\tau'_{m,n} &= \pi i \Bigl[+ \frac{1}{\hat p}\Bigl(m-(\nu'-\nu)\Bigr) 
+ \frac{1}{\hat q}\Bigl(n-(\nu'+\nu)\Bigr)\Bigr],
\end{align}
and the corresponding points $(\zeta_{m,n},Q_{m,n})$ are expressed as 
\begin{align}
\zeta_{m,n} &= (-1)^{m}\lambda^{\hat p} 
\cos \Bigl[
\pi \frac{(\hat q-\hat p) (\nu'+\nu) + n \hat p}{\hat q} 
\Bigr], \nn\\
Q_{m,n} &= (-1)^{n}\lambda^{\hat q} 
\cos \Bigl[
\pi \frac{(\hat q-\hat p) (\nu'-\nu) - m \hat q}{\hat p} 
\Bigr].  \label{SolOfSingularPoints}
\end{align}
Among these points, the branch points for $Q(\zeta)$ (or $\zeta(Q)$) appear when 
\begin{align}
\nu'+ \nu \in \mathbb Z,\qquad (\text{or}\ \nu'-\nu \in \mathbb Z). \label{chargeBranch}
\end{align}

\subsubsection{The $\mathbb Z_{k}$ charges of D branes and singular points}
For later convenience, 
here we introduce the $\mathbb Z_{k}$ charge according to 
the original $\mathbb Z_k$ charge conjugation 
of the $k$-cut matrix models, even though our critical points 
do not preserve the $\mathbb Z_k$ symmetry. 
Each eigenvalue of the operators $\bP$ and $\bQ$ 
corresponds to each FZZT brane with a different charge \cite{fi1}.%
\footnote{This is reviewed in Appendix \ref{ZkChargeFreeFermion}}
Therefore, the phase $\nu$ defines the $\mathbb Z_k$ charge of FZZT branes as 
\begin{align}
\exp[2\pi i \nu]. 
\end{align}
According to \cite{fi1}, the ZZ branes also have $\mathbb Z_{\hat k}$ charges. A natural definition here 
is that branch points \eq{chargeBranch} do not have this charge, which means that 
\begin{align}
\exp\bigl[\pm 2\pi i (\nu'+\nu)\bigr]. \label{ZkChargeZZbrane}
\end{align}
The plus-minus sign means that there is a pair of ZZ brane and its charge conjugate at each singular point. 

Although the critical points break the $\mathbb Z_k$ symmetry, this system still admits 
the following $\mathbb Z_2$ charge conjugation, 
\begin{align}
\nu \to - \nu. 
\end{align}
This $\mathbb Z_2$ charge conjugation is understood 
as a result of the hermiticity \eq{HermiticityPotentials}, 
and the D-branes of $\nu=0,\pm 1/2$ are self-charge conjugate D-branes. 

\subsubsection{Counting of singular points}
In this section, we count the number of singular points. 
First of all, one should resolve the basic periodicity of the parameters $\tau$ and $\tau'$ 
in Eq.~\eq{SingularBranchesCond2}. Since the integers $(\hat p,\hat q)$ are coprime, 
the paramters are identified as 
\begin{align}
\tau \sim \tau + 2\pi i,\qquad \tau' \sim \tau' + 2\pi i, \label{PeriodTau}
\end{align}
which can be rephrased in terms of the parametrization $(m,n)$ as follows: 
\begin{align}
(m,n) \sim (m+ \hat p,n+\hat q) \sim (m+\hat p, n-\hat q). 
\end{align}
Therefore, the parameterization $(m,n)$ can be restricted to the following fundamental domain, for instance, 
\begin{align}
\quad 0 \leq m\leq \hat p-1,\qquad 0\leq n\leq 2\hat q-1. \label{FundMN}
\end{align}
Secondly, one has to avoid further double counting due to 
some accidental coincidence of the points $(\zeta_{m,n},Q_{m,n}) = (\zeta_{m',n'},Q_{m',n'})$
in the expression Eq.~\eq{SolOfSingularPoints}. This happens
if $(\hat q-\hat p)$ and $(\nu,\nu')$ satisfy either of the following two conditions: 
\begin{itemize}
\item The first condition is related to $\cosh ( - x ) = \cosh(x)$:
\begin{align}
(\hat q-\hat p)(\nu'- \nu) = l_- \in \mathbb Z \qquad \text{or}\qquad 
(\hat q-\hat p)(\nu'+ \nu) = l_+ \in \mathbb Z. \label{AccidentalSym1}
\end{align}
In these cases, the expression \eq{SolOfSingularPoints} turns out to be symmetric 
under the following transformations, respectively:
\begin{align}
(m,n) \to (-(m-m_-)+m_-,n) \quad \text{or}\quad 
(m,n) \to (m,-(n-n_+)+n_+), \label{AccidentalSymTrans1}
\end{align}
Here $m_-$ and $n_+$ are integers which satisfy 
\begin{align}
\hat q m_\pm \mp \hat p n_\pm = l_\pm =(\hat q- \hat p)(\nu'\pm \nu) \in \mathbb Z,  
\end{align}
with some proper integers of $n_-$ and $m_+$. In the fundamental domain \eq{FundMN}, 
there is a unique $m_-$ and are two different $n_+$. 
\item The second condition is related to $\cosh (-x + \pi i/2) = -\cosh (x + \pi i/2)$:
\begin{align}
(\hat q-\hat p)(\nu'- \nu) = l_- \in \frac{2 \mathbb Z +1}{2}\quad \text{and} \quad 
(\hat q-\hat p)(\nu'+ \nu) = l_+ \in \frac{2 \mathbb Z +1}{2}. \label{AccidentalSym2}
\end{align}
In this case,
the expression \eq{SolOfSingularPoints} turns out to be symmetric under the following transformation:
\begin{align}
(m,n) \to (-(m-m_-)+m_-,-(n-n_+)+n_+).  \label{AccidentalSymTrans2}
\end{align}
Here $m_-$ and $n_+$ are half integers which satisfy
\begin{align}
\hat q m_\pm \mp \hat p n_\pm = l_\pm = (\hat q- \hat p)(\nu'\pm \nu)  \in \frac{2 \mathbb Z +1}{2}, 
\end{align}
with proper half integers of $m_+$ and $n_-$. 
\end{itemize}
Taking into account these facts, we show the number of singular points in the following:

\paragraph{\underline{(i) The singular points among the same curves}}
\ \vspace{0.4cm} 
\\
The singular points among the same curves happens when the charges $\nu$ and $\nu'$ satisfy 
\begin{align}
\text{neutral ZZ branes:} \quad \nu' +\nu \, \in\,  \mathbb Z,\qquad
\text{charged ZZ branes:} \quad \nu' -\nu \, \in\,  \mathbb Z. 
\end{align}
In these cases, the points \eq{SolOfSingularPoints} also include non-singular points like 
the branch points which happen if 
the pair $(\tau_{m,n}, \tau'_{m,n})$ satisfies the following condition: 
\begin{align}
\tau'_{m,n} +  \tau_{m,n} \,\in\, 2\pi i \mathbb Z\qquad  (\text{when}\ \ \nu' + \nu \,\in\, \mathbb Z) 
\qquad &\Leftrightarrow \qquad n -(\nu'+\nu)\,\in\, \hat q \,\mathbb Z, \label{BranchPointConditions2}\\
\tau'_{m,n} -  \tau_{m,n} \,\in\, 2\pi i \mathbb Z\qquad  (\text{when}\ \ \nu' - \nu \,\in\, \mathbb Z) 
\qquad &\Leftrightarrow \qquad m -(\nu'-\nu)\,\in\, \hat p \,\mathbb Z. \label{BranchPointConditions1}
\end{align}
Therefore, we have to exclude these cases from the counting. 

If $\nu' = \nu = 0$ or $1/2$, then this can satisfy both 
of the conditions \eq{BranchPointConditions1} and \eq{BranchPointConditions2}. 
In this case, we can say that there is no charged ZZ branes because this FZZT brane is self dual. 
By taking into account the accidental 
symmetry of Eqs.~\eq{AccidentalSym1} and \eq{AccidentalSym2}, one can count the number of ZZ branes 
which is given as 
\begin{align}
{\nu' - \nu,\, \nu'+\nu \in \mathbb Z:}&\qquad \frac{(\hat p-1)(\hat q-1)}{2}. 
\end{align}
The other cases satisfy part of the condition \eq{BranchPointConditions1} and \eq{BranchPointConditions2}. 
The cases of neutral ZZ branes ($\nu' + \nu \in \mathbb Z$) are given as 
\begin{align}
{\nu' + \nu \in \mathbb Z,\,\, 2(\hat q-\hat p)\nu,\, \nu'-\nu \notin \mathbb Z:}
&\qquad \hat p (\hat q-1), \nn\\
{\nu' + \nu, \,2(\hat q-\hat p)\nu \in \mathbb Z,\,\, \nu'-\nu  \notin \mathbb Z:}
&\qquad \frac{(\hat p+1)(\hat q -1)}{2}. 
\end{align}
The cases of charged ZZ branes ($\nu' - \nu \in \mathbb Z$) are given as 
\begin{align}
{\nu' - \nu \in \mathbb Z,\,\, 2(\hat q-\hat p)\nu,\,\nu'+\nu \notin \mathbb Z:}
&\qquad (\hat p-1)\hat q, \nn\\
{\nu' - \nu, \,2(\hat q-\hat p)\nu \in \mathbb Z,\,\, \nu'+\nu  \notin \mathbb Z:}
&\qquad \frac{(\hat p-1)(\hat q +1)}{2}. 
\end{align}

\paragraph{\underline{(ii) The singular points between different curves}}
\ \vspace{0.4cm} 
\\
In these cases, all the ZZ branes are charged, 
and it is convenient to write the charges as follows:
\begin{align}
(\nu' -\nu,\nu'+\nu) = \bigl(\frac{u}{rt},\frac{v}{st}\bigr) 
\qquad \Leftrightarrow \qquad
(\nu,\nu') = \bigl(\frac{vr-us}{2rs t},\frac{su+rv}{2rs t}\bigr). 
\end{align}
Here $\{r,s,t\}$ are positive non-zero integers, $\{u,v\}$ are non-zero integers 
and $(r,s),(rt,u)$ and $(st,v)$ are pairs of coprime integers. 
In these cases, there is no branch point and the only thing one needs to take into account is 
the accidental coincidence \eq{AccidentalSym1} and \eq{AccidentalSym2} among $2\hat p\hat q$ 
points $(m,n)$. The numbers are following:
\begin{align}
{(\hat q-\hat p)\notin rt\, \mathbb Z \cup st\, \mathbb Z:}
&\qquad 2 \hat p \hat q, \nn\\
{(\hat q-\hat p)\in rt\, \mathbb Z \setminus  rst\, \mathbb Z:}
&\qquad (\hat p+1) \hat q, \nn\\
{(\hat q-\hat p)\in st\, \mathbb Z \setminus  rst\, \mathbb Z:}
&\qquad \hat p(\hat q+1), \nn\\
{(\hat q-\hat p)\in rst\, \mathbb Z:}
&\qquad \frac{(\hat p+1)(\hat q+1)}{2}, 
\end{align}
and 
\begin{align}
{(\hat q-\hat p)\in \,rs\frac{t}{2}\,(2 \mathbb Z+1)\,\wedge\, (r-s) \in 2\mathbb Z:}
\qquad  \hat p \hat q. 
\end{align}
Note that the last case should be $u,v \in 2\mathbb Z+1$ which is automatic because $(tr,u)$ and $(ts,v)$ are coprime integers.

\section{The FZZT- and ZZ-brane amplitudes \label{SectionFZZTandZZamp}}

In this section, we calculate the FZZT- and ZZ-brane amplitudes from our macroscopic loop amplitudes. 
We first review the relations between these brane amplitudes
and the multi-cut matrix models before giving detailed derivations.

\subsection{The FZZT-brane amplitudes}
\subsubsection{The $\mathbb Z_k$ charged FZZT-branes from the matrix models}

Originally the FZZT brane is related to the orthonormal polynomials \cite{FZZT} as
\begin{align}
\alpha_n(x) \sim \vev{\det \bigl(x-X\bigr)} 
= \exp\bigl[\tr \ln (x-X)\bigr] 
= \exp\bigl[\frac{1}{g_{\rm str}} D_{\rm FZZT}(\zeta) + \cdots\bigr]. \label{FZZTact}
\end{align}
A new feature of the multi-cut matrix models is the $\mathbb Z_k$ charge 
with respect to the charge conjugation \eq{ZkChargeConjInt}. 
Although the $\mathbb Z_k$ charges are clearly formulated in the free-fermion system 
\cite{fi1} (reviewed in Appendix \ref{ZkChargeFreeFermion}), 
description of the $\mathbb Z_k$ charge within matrix models is not trivial. 

A possible answer was provided in section 
\ref{MultiCutGeometryStokesSection}. 
That is, the asymptotic behavior of the orthonormal polynomial \eq{FZZTact} 
depend on the regime of $\zeta$ (see Eq.~\eq{StokesBakerAnaly}). 
Here we claim that {\em the $\mathbb Z_k$ charge $e^{2\pi i \nu_j}$ of FZZT brane 
corresponds to the asymptotic regime $\zeta \to \infty \times e^{2\pi i \nu_j}$ of 
the orthonormal polynomials \eq{FZZTact}}. Consequently, there is only one kind of 
FZZT brane in this theory, and the $\mathbb Z_k$ charge of the FZZT brane 
expresses the position of the FZZT brane in the $\zeta$ plane. 
Our description is natural because originally there is only one kind of matrix-model eigenvalue $x$ 
in the multi-cut matrix models, and also because this consideration 
is consistent with the original charge conjugation \eq{ZkChargeConjInt}. 
This also gives a support for our conjecture \eq{ConnectionRelation} given in section 
\ref{MultiCutGeometryStokesSection}. 

Interestingly, since each asymptotic regime 
corresponds to a weak-coupling regime of string theory, 
this argument indicates that 
the Ramond-Ramond charge of FZZT brane is a concept 
which appears as a perturbative string artifact. 
In this description, the FZZT-brane charge is the label of asymptotic regimes in spacetime, 
and therefore resembles a spacetime coordinate. As noted in Introduction, 
this description bears strong resemblance to the Kaluza-Klein 
compactification of gravity theory \cite{KaluzaKlein}: 
compactification of a dimension results in $U(1)$ gauge fields. 
This idea will be elaborated in section \ref{SecConclusion}. 

According to this proposal, the $\mathbb Z_k$ charged FZZT-brane partition function 
$\mathcal Z_{\rm FZZT}^{(j)}(t;\zeta)$ 
with a charge $e^{2\pi i \nu_j}$ 
is the orthonormal polynomials \eq{FZZTact} 
or the Baker-Akhiezer function $\Psi(t;\zeta)$ in a particular perturbative regime:%
\footnote{The Baker-Akhiezer functions are now vector valued funtions. 
Therefore, they have $k$ different functions as components. 
However it is known that they are related 
to the same FZZT brane with different background (flux) \cite{SeSh2}. Therefore, 
the leading disk amplitudes (or FZZT-brane action) of these components are the same. }
\begin{align}
\mathcal Z_{\rm FZZT}^{(j)}(t;\zeta) \equiv \Psi(t; e^{2\pi i\nu_j}\zeta)
= \exp\bigl[\frac{1}{g_{\rm str}} D_{\rm FZZT}^{(j)}(\zeta;t) + \cdots\bigr]\qquad 
(\zeta\to \infty\times e^{2\pi i \nu_0}). 
\label{FZZTact2}
\end{align}
Therefore, the FZZT-brane disk amplitude is  
\begin{align}
D_{\rm FZZT}^{(j)}(\zeta,t) &= \lambda^{\hat p+\hat q}\int^{\zeta} 
\, d\Pi^{(j)}(z)
\, \Xi^{(j)}(z),\qquad \zeta = e^{-2\pi i\nu_j} \lambda^{\hat p}\Pi^{(j)}(z). 
\end{align}
See Eq.~\eq{DimensionSeparation} 
for the definitions of the functions $\Pi^{(j)}(z)$ and $\Xi^{(j)}(z)$. 

\subsubsection{The FZZT-brane disk amplitudes}
With noting the relation of cosmological constant 
$\mu = \lambda^{2\hat p}\,(=t^{\frac{2\hat p}{\hat p+\hat q-1}})$, 
the FZZT-brane disk amplitudes in the fractional superstring theory 
are calculated as
\begin{align}
D_{\rm FZZT}^{(j)}(\zeta,\mu) &= \lambda^{\hat p+\hat q}\int^{\zeta} 
\, d\Pi^{(j)}(z)
\, \Xi^{(j)}(z) \nn\\
&=\hat p\, (\sqrt{\mu})^{\frac{\hat q}{\hat p}+1} 
\int^{\tau} d\tau' \sinh(\hat p \tau' +2\pi i \nu_j) \, \cosh(\hat q \tau' + 2\pi i \nu_j) \nn\\
&= \frac{\hat p }{2} \, (\sqrt{\mu})^{\frac{\hat q}{\hat p}+1} 
\Bigl[
\frac{\cosh\bigl((\hat q +\hat p)\tau +4\pi i \nu_j\bigr) }{\hat q+ \hat p}
-
\frac{\cosh\bigl((\hat q-\hat p)\tau\bigr) }{\hat q- \hat p}
\Bigr],  \label{FZZTamp}
\end{align}
with $\zeta = e^{-2\pi i \nu_j}\sqrt{\mu} \cosh\bigl(\hat p \tau + 2\pi i \nu_j\bigr)$. 
The modular S-matrices on the Liouville side are directly related to one-point function of (bulk) cosmological constant 
operator \cite{SeSh} which can be easily calculated from the above expression as%
\footnote{Note that we take a derivative of $\mu$ with $\zeta$ fixed. 
Therefore, one needs to take into account 
Eq.~\eq{ZetaSigmaB}. }
\begin{align}
\del_\mu D_{\rm FZZT}^{(j)}(\zeta,\mu) \Bigr|_\zeta
&= \frac{\hat q}{2(\hat p-\hat q)} 
\bigl(\sqrt{\mu}\bigr)^{\frac{\hat q}{\hat p}-1} \cosh (\hat q-\hat p)\tau, \nn\\
&= \frac{1}{2(b^2-1)} 
\bigl(\sqrt{\mu}\bigr)^{\frac{1}{b^2}-1} \cosh \Bigl[\pi \Bigl(b-\frac{1}{b}\Bigr)\sigma\Bigr],
\end{align}
where we introduce the CFT parameters, 
the Liouville coupling $b$ and open channel momentum $\sigma$:
\begin{align}
b= \sqrt{\frac{\hat p}{\hat q}},\qquad 
\sigma = \frac{1}{\pi} \sqrt{\hat p \hat q}\,  \tau. \label{LiouvilleParam}
\end{align}
The expression here is the same as bosonic and superstring Liouville calculation (See \cite{SeSh}). 
This means that the modular S-matrices for non-degenerate representations might be 
the same as bosonic or superstring cases. The difference comes from the relation to 
the boundary cosmological constant:
\begin{align}
\zeta &= e^{-2\pi i \nu_j} \sqrt{\mu} \cosh\bigl(\hat p \tau + 2\pi i \nu_j\bigr) 
= e^{-2\pi i \nu_j} \sqrt{\mu} \cosh \bigl(\pi b \sigma  + 2\pi i \nu_j\bigr).  
\label{ZetaSigmaB}
\end{align}
This information would be helpful for the boundary conformal bootstrap analysis 
\cite{FZZT,fuku-hoso} of fractional super Liouville field theory. 

\subsection{The ZZ-brane amplitudes}

\subsubsection{The charged ZZ-brane from the matrix models \label{ZZfromMM}}
The instanton disk amplitudes were first studied in 
\cite{GinspargZinnJustin,EynardZinnJustin}, and then realized in the 
free-fermion analysis \cite{fy12,fy3}. 
The appearance of the ZZ-brane disk amplitudes in the two-matrix models 
was systematically studied in \cite{KazakovKostov}, and then 
also in the free-fermion analysis \cite{fis,fim} including 
two-cut cases \cite{fi1} which reproduce the results from the 
super-Liouville calculations \cite{SeSh}. 
According to \cite{fi1}, 
the definition of the amplitudes in the multi-component cases is simply given as \eq{DefZZ}. 
Realization of charged ZZ branes in the multi-cut matrix models can also be understood 
by taking the differences of FZZT branes. Here we make a parallel comparison 
with \cite{KazakovKostov} to see that a straightforward generalization, Eq.~\eq{DefZZ}, 
of the amplitude calculations can be justified in the cases of multi-cut two-matrix models. 

According to the saddle point method of the matrix models \cite{David} (one-matrix) 
and \cite{KazakovKostov} (two-matrix), the partition function of the two-matrix model 
is given as
\begin{align}
\mathcal Z &= \int dX dY e^{-N \tr w(X,Y)} \nn\\
&= \int dx dy \, e^{-N w(x,y)}\, 
\Bigl< \det \bigl(x-X_{(N-1)}\bigr) \det \bigl(y-Y_{(N-1)}\bigr)\Bigr>_{(N-1)} \nn\\
&\equiv  \int dx dy\, e^{-N S_{\rm eff}(x,y)}.
\end{align}
In this mean field approximation (in $N\to \infty$) \cite{David}, 
one considers that the rest $N-1$ eigenvalues are in semi-classical configuration,%
\footnote{
In the case of the multi-cut matrix models, 
it is known that we need to fix the filling fraction and sum over these configulations 
\cite{MultiCutUniversality}. 
However, since our main purpose is to justify the ZZ-brane disk amplitudes \eq{DefZZ}, 
we make an intuitive argument below. }
and the expectation values are factorized:
\begin{align}
\Bigl< \det \bigl(x-X\bigr) \det \bigl(y-Y\bigr)\Bigr> 
= \exp\Bigl[N\vev{\dfrac{1}{N}\tr \ln(x-X)}+N\vev{\dfrac{1}{N}\tr \ln(y-Y)} +O(1)\Bigr],
\end{align}
and the effective action $S_{\rm eff}(x,y)$ is expressed as follows:
\begin{align}
S_{\rm eff}(x,y)= \Phi (x) + \tilde \Phi (y) 
-xy,
\end{align}
with 
\begin{align}
\Phi(x) = V_1(x)-\Bigl<\frac{1}{N} \tr \ln \bigl(x - X)\Bigr>,\qquad 
\tilde \Phi(x) = V_2(y)-\Bigl<\frac{1}{N} \tr \ln \bigl(y - Y)\Bigr>. \label{MacroInInstAction}
\end{align}

As we have discussed in section \ref{MultiCutGeometryStokesSection}, 
these effective potentials (or macroscopic loop amplitudes) 
$\Phi(x)$ and $\tilde \Phi(y)$ take different asymptotic values (as \eq{SolQofJ}) 
according to the position of $x$ or $y$ as in Eq.~\eq{ConnectionRelation}. 
Therefore, the effective potentials (outside the cuts)%
\footnote{Note that this is the expression outside the cuts on physical sheet. 
One can consider the expression on the cuts, and it should be a superposition of 
the macroscopic loop amplitudes on two sides of the cuts 
(e.g.~$Q^{(j)}(\zeta)+Q^{(j+1)}(\zeta)$). 
However, it is not so important here 
because instanton actions always appear outside of the cuts. }
are given as
\begin{align}
N \Phi(x) &= N \Phi^{(j)}(x) 
\equiv \frac{-\lambda^{\hat p+\hat q}}{g_{str}}
\int^{\zeta} \Xi^{(j)}(z)\, d \Pi^{(j)}(z),\qquad 
\zeta = \lambda^{\hat p}\Pi^{(j)}(z),\nn\\
N \tilde \Phi(y) &= N \tilde \Phi^{(j')}(y) 
\equiv \frac{-\lambda^{\hat p+\hat q}}{g_{str}}
\int^{\tilde\zeta } \Pi^{(j')}(z')\, d \Xi^{(j')}(z'), 
\qquad \tilde \zeta = - \lambda^{\hat q}\Xi^{(j')}(z'), 
\end{align}
on each asymptotic regime ($j,j' = 1,2,\cdots,k$) of $z\to \infty$. 
We should note Eq.~\eq{MacroLaxRelation}, especially the relation 
between dual macroscopic loop amplitudes $\tilde Q(\tilde \zeta)$ and Lax operators. 
The eigenvalue $x$ and $\zeta$ ($y$ and $\tilde \zeta$) are related as 
Eq.~\eq{RelXandZeta}. 
Therefore, on each sector the effective action is given as 
\begin{align}
S_{\rm eff}(x,y) = S_{\rm eff}^{(j,j')}(x,y)= \Phi^{(j)} (x) + \tilde \Phi^{(j')} (y) 
-xy. \label{EffectiveActionSectors}
\end{align}
This is the origin of the $\mathbb Z_k$ charged ZZ branes in the multi-cut matrix models. 

\begin{figure}[htbp]
 \begin{center}
  \includegraphics[scale=0.9]{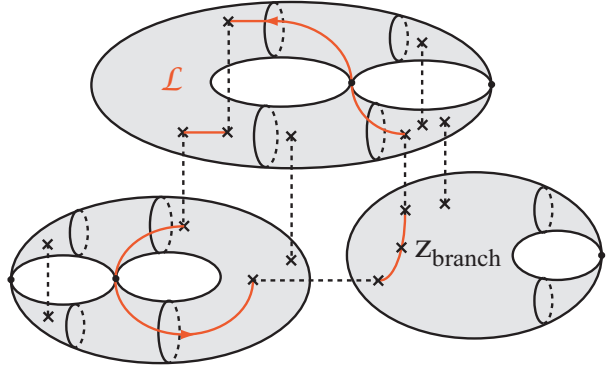}
 \caption{\footnotesize 
Typical curves and a typical instanton loop $\mathcal L$ in the multi-cut two-matrix models. 
Non-trivial connections among the irreducible curves (drawn by doted lines) are also singular points and are 
related to charged ZZ branes (also see \cite{SeSh}). 
Each Instanton loop forms a single topological loop and should include the starting point 
$z_{\rm branch}$ which defines the perturbative action.  }
 \label{instanton}
  \end{center}
\end{figure}

As in \cite{KazakovKostov}, the saddle point equations, 
$\del_x S_{\rm eff}^{(j,j')}(x,y) = \del_y S_{\rm eff}^{(j,j')}(x,y) = 0$, give 
possible saddle points $(x_*,y_*)$, 
\begin{align}
y_* = \frac{\del\Phi{}^{(j)}}{\del x} (x_*) \qquad &\Leftrightarrow\qquad
(x_*,y_*) = (a^{\frac{\hat p}{2}} \lambda^{\hat p}\Pi^{(j)}(z), 
-a^{\frac{\hat q}{2}}\lambda^{\hat q} \Xi^{(j)}(z)),\nn\\
x_* = \frac{\del\tilde \Phi{}^{(j')}}{\del y} (y_*) \qquad &\Leftrightarrow\qquad
(x_*,y_*) = (a^{\frac{\hat p}{2}} \lambda^{\hat p} \Pi^{(j')}(z'), 
-a^{\frac{\hat q}{2}}\lambda^{\hat q} \Xi^{(j')}(z')),
\end{align}
which are the intersection conditions of the curves 
(which have been studied in section \ref{AlgSingularPointsSection}):
\begin{align}
(x_*,y_*) =(x_{m,n}^{(j,j')},y_{m,n}^{(j,j')})
\equiv (a^{\frac{\hat p}{2}}\zeta_{m,n}^{(j,j')},
-a^{\frac{\hat q}{2}}Q_{m,n}^{(j,j')}). 
\end{align}
Among them, usually the branch points on physical sheet are most stable saddle points%
\footnote{If there is no branch point on the physical sheet, one needs to take one of the singular points. } and 
this is the definition of perturbative action \cite{KazakovKostov}. Therefore, instanton actions 
are evaluated at the saddle points of singular points. By using partial integration, 
the action \eq{EffectiveActionSectors} becomes 
\begin{align}
NS_{\rm eff}(x_{m,n}^{(j,j')},y_{m,n}^{(j,j')}) 
=& \frac{-\lambda^{\hat p+\hat q}}{g_{\rm str}}\int_{{\rm cut off}}^{z_{\rm branch}}
\Bigl(d \Pi^{(j)}(z) \, \Xi^{(j)}(z) - d \Pi^{(j')}(z') \, \Xi^{(j')}(z') \Bigr)+\nn\\
&+ \frac{-\lambda^{\hat p+\hat q}}{g_{\rm str}}
\Bigl(\int_{z_{\rm branch}}^{z_{m,n}^{(j,j')}} d\Pi^{(j)}(z) \, \Xi^{(j)}(z) 
- \int_{z_{\rm branch}}^{z_{-m,n}^{(j',j)}} d\Pi^{(j')}(z) \, \Xi^{(j')}(z)\Bigr).
\end{align}
The first term is the perturbative action which is usually non-universal and 
the second term is the instanton action \cite{KazakovKostov}. 
As one can see that the contour of the instanton action forms a closed loop $\mathcal L$ 
in the whole curve $F(\zeta,Q)=0$
(see Fig.~\ref{instanton}), 
\begin{align}
S_{\rm inst.}[\mathcal L] = \frac{-\lambda^{\hat p+\hat q}}{g_{\rm str}}\oint_{\mathcal L} 
d\Pi^{(j)}(z) \, \Xi^{(j)}(z),\qquad 
z_{\rm branch} \in \mathcal L,
\end{align}
which can be decomposed into a sum of elemental pieces, i.e.~ZZ-brane disk amplitudes (see \eq{DefZZ}), 
and is related to the vacuum energy of each irreducible curve in the weak-coupling limit. 

\subsubsection{The ZZ-brane disk amplitudes}

The ZZ-brane amplitudes are now defined as the following difference between 
the corresponding pair of FZZT brane amplitudes in Eq.~\eq{FZZTamp} on the singular points:
\begin{align}
&D_{\rm ZZ}^{(j,j')}(m,n;\mu)
= D_{\rm FZZT}^{(j)}(\zeta (\tau_{m,n}^{(j,j')}),\mu) 
-D_{\rm FZZT}^{(j')}(\zeta (\tau'{}_{m,n}^{(j,j')}),\mu) \nn\\
&\qquad = 
\frac{2\hat p\hat q\, (\sqrt{\mu})^{\frac{\hat q}{\hat p}+1} }{\hat p^2-\hat q^2} 
\sin\Bigl[ \frac{(\hat q -\hat p) \bigl(n-(\nu_{j'}+\nu_j)\bigr)}{\hat q} \pi \Bigr]
\sin\Bigl[ \frac{(\hat q -\hat p)\bigl(m-(\nu_{j'}-\nu_j)\bigr)}{\hat p} \pi \Bigr].  \label{DefZZ}
\end{align}
The derivative in the bulk cosmological constant $\mu$ is just given as
\begin{align}
&\del_\mu D_{\rm ZZ}^{(j,j')}(m,n;\mu) = \nn\\
&\qquad =
\frac{(\sqrt{\mu})^{\frac{1}{b^2}-1} }{b^2-1} 
\sin\Bigl[ (1 -b^2) \bigl(n-(\nu_{j'}+\nu_j)\bigr) \pi \Bigr]
\sin\Bigl[ (\dfrac{1}{b^{2}} -1)\bigl(m-(\nu_{j'}-\nu_j)\bigr) \pi \Bigr]. 
\end{align}
Of course, this expression coincides with the one-cut calculation 
\cite{GinspargZinnJustin,EynardZinnJustin}\cite{KazakovKostov} 
and the two-cut calculations 
in terms of the free fermion formulation \cite{fi1} and in the super-Liouville theory \cite{SeSh}. 

\section{Conclusion and discussion \label{SecConclusion}}

\subsection{Fractional-superstring amplitudes and Liouville theory}
In this paper, we have studied macroscopic loop amplitudes 
of $(\hat p,\hat q)$ minimal $k$-fractional superstring theory
within the $k$-cut two-matrix models in the fractional-superstring critical points. 
Even though the system breaks the $\mathbb Z_k$ symmetry of the $k$-cut two-matrix models, 
the matrix models are still solvable in all of the critical points labeled by $(\hat p,\hat q;k)$ 
and the amplitudes can be expressed as a hyperbolic cosine or sine function. 
Although there is a difference between $\omega^{1/2}$-rotated critical points and real-potential critical points, 
the main structure of the solutions is the same and is simply given as 
\begin{align}
\text{cosh solutions:}&\qquad 
\zeta = \sqrt{\mu} \cosh(\hat p \tau + 2\pi i \nu_j),\qquad 
Q^{(j)}= \mu^{\frac{\hat q}{2\hat p}}\cosh(\hat q \tau + 2\pi i \nu_j),  \nn\\
\text{sinh solutions:}&\qquad 
\zeta = \sqrt{\mu} \sinh(\hat p \tau + 2\pi i \nu_j),\qquad 
Q^{(j)}= \mu^{\frac{\hat q}{2\hat p}}\sinh(\hat q \tau + 2\pi i \nu_j), \label{SolutionsConclusion}
\end{align}
with $\nu_j=(j-1)/k$ ($\omega^{1/2}$-rotated potentials) or 
$\nu_j=(2j-1)/2k$ (real potentials). 

Since our investigation is a search for possible asymptotic solutions of the fractional-superstring critical points 
in the multi-cut matrix models, non-perturbative analysis and Liouville theory calculations are pending, 
and we hope to address these points in future investigations. 
Possible issues related to the current work are listed in order:
\begin{itemize}
\item There is only one type of solutions in the odd-cut models. According to 
the analogy to bosonic strings and type 0 superstrings \cite{NewHat}, 
this indicates that there is a bound in the cosmological constant, say $\mu >0$. 
This implies the following: The cosmological constant $\mu$ would be related 
to the ``bosonic'' cosmological constant $\mu_{\rm bos}$ as $\mu_{\rm bos} = \mu^k$, 
which plays a role of suppressing large-area random surfaces $(A\to \infty)$
in the Liouville theory partition function $\mathcal F_h$ of topology $h$:
\begin{align}
\mathcal F_h \sim \sum_{\text{surfaces of topology $h$}} e^{-\mu_{\rm bos} \times A  + \cdots}, \qquad 
(\mu_{\rm bos} = \mu^k).
\end{align} 
In order to make this integral well-defined, 
there is a constraint on the cosmological constant, $\mu_{\rm bos}=\mu^k>0$. 
On the other hand when $k$ is even, it is natural to say that the cosh and sinh solutions would be exchanged 
with each other depending on the sign changing of the cosmological constant $\mu$: $\mu<0 \to 0<\mu$. 
These points should be checked in some actual formulation of fractional super-Liouville theory 
and/or non-perturbative analysis of string equations. 
\item We have obtained two types of critical points: $\omega^{1/2}$-rotated potentials and real potentials. 
Our results show that they are distinct when $k\in 4\mathbb Z$. 
From the worldsheet point of view, the specialty of the $k \in 4\mathbb Z$ cases 
is the appearance of an extra bosonic (local) current.%
\footnote{When $k$ is even, among the parafermion fields $\{\psi_j(z)\}_{j=1}^k$, 
there is a special energy operator $\epsilon_{k/2}(z)=\psi_{k/2}(z)$, 
whose conformal dimension is $k/4$. Therefore, this is a local field. 
This becomes a bosonic field when $k\in 4\mathbb Z$.}
Therefore, it is interesting if there is a connection between this fact and our results. 
\item As we have argued in section \ref{MultiCutGeometryStokesSection}, 
the highly reducibility of the algebraic equation indicates
perturbative decoupling of several operators. 
This should be observed also in the Liouville 
calculations, as an extension of the calculation given in \cite{DiFKutasov}. 
A succussful varification of this point will provide a non-trivial check of the correspondence between the multi-cut matrix models 
and fractional super-Liouville theory \cite{irie2}. 
\end{itemize}

\subsection{The multi-cut matrix models and non-critical M theory}

In this section, we give a discussion about a possible connection to the non-critical M theory 
proposed by P.~Ho\v rava and C.~A.~Keeler \cite{NonCriticalMTheory}. 

Basic proposal given by \cite{NonCriticalMTheory} is 
{\em ``to identify the extra dimension of noncritical M-theory with the angular variable on the eigenvalue 
plane''}, and in order to realize this scenario, 
they introduced {\em ``the double-scaled nonrelativistic Fermi system in upside-down harmonic potential.''}
In their proposal, the angular coordinate is identified so that the Ramond-Ramond flux of D0-brane 
in 0A background can be understood as Kaluza-Klein momentum 
of the Kaluza-Klein reduction along this angular direction. In this three (or 2+1) dimensional theory, 
type 0A and type 0B superstring theory are also shown to be included in its Hilbert space \cite{NonCriticalMTheory}. 

On the other hand, in the multi-cut matrix models, there is the $\mathbb Z_k$ charged D-branes 
and the $\mathbb Z_k$ charge conjugation is the rotation in the eigenvalue space, or a
translation of the Stoke labeling $\nu$:
\begin{align}
\nu \to \nu+ a, \qquad \nu \sim \nu + \mathbb Z, 
\end{align}
which moves one Stokes sector to another. 
Therefore, the fractional superstring theory is also naturally understood as a Kaluza-Klein reduction 
about this non-perturbative direction $\nu$ in a three-dimensional gravity theory. 
We here propose that this $U(1)$ direction can be 
a possible candidate of {\em the angular direction} in the non-critical M theory. 

First of all, let us mention the $\hat c=1$ limit of our fractional-superstring amplitudes in order to 
compare the fractional-superstring amplitudes to the spectral curve of the HK non-critical M theory.

\subsubsection{The $\hat c=1$ limit and the Ho\v rava-Keeler non-critical M theory \label{NCMtherySection}}

The $\hat c=1$ limit is simply defined by the following limit of indices $(\hat p,\hat q)$ \cite{KazakovKostov}:
\begin{align}
\hat p,\hat q \to \infty, \qquad b^2 \equiv \hat p/\hat q \to 1. 
\end{align}
This limit is interesting because it can relate the system 
to the $\hat c =1$ non-critical fractional superstring theory%
\footnote{The definition of $\hat c$ in $k$-fractional superstring theory is given as 
$\hat c \equiv \dfrac{k+2}{3k} c$. In this notation, $\hat c=1$ means that spacetime dimension is two 
(i.e.~Liouville + matter). } 
or {\em multi-cut matrix quantum mechanics} \cite{irie2}. 

\begin{figure}[htbp]
 \begin{center}
  \includegraphics[scale=0.6]{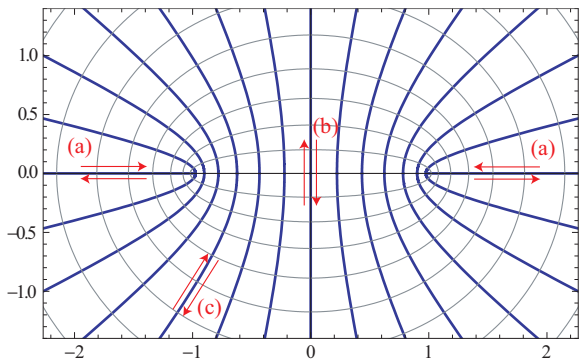}
 \caption{\footnotesize 
The complex $x$ space trajectories in $k=28$ cut $\hat c=1$ fractional strings (cosh solution) 
with an interpretation of real $\sigma$ as proper time: 
$x=\cosh(\pi \sigma + 2\pi i \nu_j)$ with $\nu_j=(j-1)/k$.  (a) On the real axis ($\nu=0,1/2$), the 
trajectories are the same as bosonic strings: an eigenvalue comes from infinity, bounces and goes back. 
(b) On the imaginary axis ($\nu=\pm1/4$), the trajectories are the same as type 0 superstrings: an eigenvalue comes from 
infinity, goes though the tip and goes to another infinity. 
(c) Generally trajectories are similar to the superstring cases but the curves are slightly bent. }
 \label{elliptic}
  \end{center}
\end{figure}

Our cosh and sinh solution are now expressed with the Liouville parameters $b$ and $\sigma$ 
in Eq.~\eq{LiouvilleParam} as 
\begin{align}
\text{cosh solutions:}& \qquad 
\left\{
\begin{array}{l}
\ds
X^{(\nu)}(\sigma) = \sqrt{\mu} \cosh (\pi b \sigma + 2\pi i \nu),\cr
\ds
Y^{(\nu)}(\sigma) = \mu^{\frac{1}{2b^2}} \cosh (\pi \sigma/b + 2\pi i \nu),
\end{array}
\right. \nn\\
\text{sinh solutions:}& \qquad 
\left\{
\begin{array}{l}
\ds
X^{(\nu)}(\sigma) = \sqrt{\mu} \sinh (\pi b \sigma + 2\pi i \nu),\cr
\ds
Y^{(\nu)}(\sigma) = \mu^{\frac{1}{2b^2}} \sinh (\pi \sigma/b + 2\pi i \nu),
\end{array}
\right.
\end{align}
One way to obtain the spectral curve of the $\hat c=1$ system is to make use of the following combination:
\begin{align}
x^{(\nu)}(\sigma) \equiv \frac{X^{(\nu)}+Y^{(\nu)}}{2},\qquad 
y^{(\nu)}(\sigma) \equiv \frac{X^{(\nu)}-Y^{(\nu)}}{b-b^{-1}}.
\end{align}
After taking the limit $b\to 1$, this results in
\begin{align}
\text{cosh solutions:}& \qquad 
\left\{
\begin{array}{l}
\ds
x^{(\nu)}(\sigma) = \sqrt{\mu} \cosh (\pi \sigma + 2\pi i \nu),\cr
\ds
y^{(\nu)}(\sigma) = \pi \sigma \sqrt{\mu} \sinh (\pi \sigma + 2\pi i \nu)
+ \frac{\ln \mu}{2} x^{(\nu)}(\sigma),
\end{array}
\right. \nn\\
\text{sinh solutions:}& \qquad 
\left\{
\begin{array}{l}
\ds
x^{(\nu)}(\sigma) = \sqrt{\mu} \sinh (\pi \sigma + 2\pi i \nu),\cr
\ds
y^{(\nu)}(\sigma) = \pi \sigma \sqrt{\mu} \cosh (\pi \sigma + 2\pi i \nu) + \frac{\ln \mu}{2} x^{(\nu)}(\sigma).
\end{array}
\right.
\end{align}
The universal eigenvalue density function $p(\tau)$ 
or the canonical momentum of coordinate $x$ is extracted from $y^{(\nu)}(\sigma)$ 
in the following way \cite{KazakovKostov}:
\begin{align}
p^{(\nu)}(\sigma) \equiv \frac{y^{(\nu)}(\sigma-i)-y^{(\nu)}(\sigma+i)}{2\pi i} 
=
\left\{
\begin{array}{ll}
\sqrt{\mu} \sinh(\pi \sigma + 2\pi i \nu) & \text{: cosh solutions}\cr
\sqrt{\mu} \cosh(\pi \sigma + 2\pi i \nu) & \text{: sinh solutions}
\end{array}
\right.
\end{align}
Therefore, the phase space is given as
\begin{align}
\text{cosh solutions:} \quad x^2- p^2 = \mu, \qquad
\text{sinh solutions:} \quad p^2- x^2 = \mu
\end{align}
In the two-cut cases, these phase spaces reproduce the spectral curve 
of type 0B superstring theory \cite{TT,NewHat} (they are related to the $\nu=0,1/2$ sectors). 
Note that, since we consider complex integral of eigenvalues in the matrix model 
(see \eq{DefinitionOfTwoMat} and \eq{MContour}), the coordinate $x$ 
are generally complex and $\sigma$ can be understood as proper time of the motion.  
The trajectories on the complex $x$ space are shown in Fig.~\ref{elliptic}.%
\footnote{These coordinate system is known as elliptic coordinate. }

Interestingly, each motion has a preferred cosmological constant,%
\footnote{We would like to thank Kazuyuki Furuuchi, Pei-Ming Ho, Hiroshi Isono and Wen-Yu Wen
 for useful discussions about this section, and especially Kazuyuki Furuuchi for this comment. } 
$\mu_\nu \equiv \mu \cos (2\pi \nu)$.
Because of the hermiticity \eq{HermiticityPotentials}, at these turning points on the real axes, 
the potential of the model is a purely real function
and this $-\mu_\nu$ can be understood as the energy level of the Fermi sea.  
Therefore, this indicates that, for example, bosonic string vacua are more stable than 
the superstring vacuum and difference of energy level for one eigenvalue is finite. 

A main difference between the multi-cut matrix models and the HK non-critical M theory is the symmetry: 
Although the spectral curve of the multi-cut matrix models 
looks the same as that of the HK non-critical M theory 
which preserve $U(1)$ symmetry of the angular direction, 
the motions in the multi-cut matrix models break the $U(1)$ symmetry. It is because 
minimal fractional superstring theory generally breaks the $\mathbb Z_k$ (i.e. $U(1)$) symmetry. 
However note that our solutions appear in the weak coupling analysis. It is not so strange to have 
non-trivial geometry even if the theory is originally defined on a flat space coordinate.%
\footnote{For example, in the free-fermion formulation \cite{fkn,fy12,fy3}, 
originally the boundary cosmological constant $\zeta$ is simply related to the flat coordinate $\lambda$ 
as $\zeta = \lambda^{1/p}$. However, $\zeta$ turns out to be on the non-trivial Riemann surfaces 
in the weak coupling limit. }

Our discussion here relies on the argument about the $\hat c=1$ limit \cite{KazakovKostov}, 
but it should be important to define the proper multi-cut matrix quantum mechanics first and then 
to derive the above consideration from the first principle. 
This also would clarify a further connection to the double-scaled fermion system. 

We next also discuss the possibility that the multi-cut matrix models themselves 
can be interpreted as non-critical M theory. 

\subsubsection{The multi-cut matrix models as non-critical M theory?}

Not only the identification of the Stokes labeling $\nu$ 
as the angular direction of the non-critical M theory, 
this theory implies an interesting way to include perturbative string theories
(i.e.~bosonic or type 0 superstring theory) as its vacua. 

A hint is in the $\hat c=1$ type 0B superstring theory \cite{TT,NewHat}: 
In this theory, there are two Fermi seas and excitations in one of the Fermi seas 
have no communication with another excitation in the other Fermi sea 
in all order string coupling perturbation theory. 
They communicate only through non-perturbative effects \cite{Klebanov}. 
Therefore, within the perturbation theory, there is no way to distingush the excitations as those of 
type 0B superstring theory or as those of bosonic string theory. 
In this sense, one can also say that type 0B superstring theory includes perturbative bosonic string theory 
as {\em a perturbatively isolated sector}. 
The case of the fractional-superstring critical points is following: 
\begin{itemize}
\item The origin of this {\em perturbative isolation} of sectors 
in the fractional superstring cases comes from the fact that 
the macroscopic loop amplitudes obtained in the critical points are 
highly reducible \eq{FactorizationCurve1} 
and factorized into $\lfloor \frac{k}{2}\rfloor$ irreducible algebraic equations. 
As it has been mentioned in Introduction (footnote \ref{PerturbativeIsolation}), 
this indicates that these irreducible algebraic curves communicate with each other only 
through non-perturbative effects. Therefore, the fractional superstring theory includes 
$\lfloor \frac{k}{2}\rfloor$ different perturbatively isolated sectors in its vacuum or 
in the weak-coupling Landscape of this string theory.%
\footnote{As an actual check of this isolation, 
one can see the calculation of annulus amplitudes in type 0 superstring theory \cite{Irie1}, 
which shows that the two-point functions of local operators vanish 
in the one-cut phase of two-cut critical points. 
Note that annulus amplitudes between macroscopic loop amplitudes 
can have non-zero contributions which however depend on the regularization 
or definition of the operators. }
\item From our results about the macroscopic loop amplitudes \eq{SolutionsConclusion}, 
one can conclude that every perturbatively isolated sector (or perturbative string theory) 
included in the $k$-cut fractional-superstring $(\hat p,\hat q)$ critical points 
is also realized as the perturbatively isolated sector 
in $k' (\in k \, \mathbb Z)$-cut fractional-superstring $(\hat p,\hat q)$ critical points. 
In particular, the fractional $(\hat p,\hat q)$ critical points 
of the infinite-cut two-matrix models includes 
all the perturbatively isolated sectors of 
$(\hat p,\hat q)$ minimal fractional superstring theory. 
Therefore, the infinite-cut two-matrix models can be understood as a {\em mother theory} of 
$(\hat p,\hat q)$ minimal fractional superstring theory, 
and the distribution of the isolated sectors can be interpreted 
as the Landscape of $(\hat p,\hat q)$ minimal fractional superstring theory.%
\footnote{The Landscape picture appearing here is ``each irreducible algebraic curve 
corresponds to perturbative strings, and whole algebraic curve corresponds to Landscape.'' 
It is known that one can define all-order 
perturbative string theory from any algebraic curves (spectral curve) 
by using loop equations \cite{EynardLoop}. From this viewpoint, 
it is natural to consider a set of irreducible algebraic curves and regard it as a string Landscape.
It is interesting to see how we can extend the story given in this paper. }
Interestingly, in this limit, the angular direction $\nu$ discussed in the previous section 
becomes continuum coordinate, and the three-dimensional theory appears. 
\item Appearance of the extra-third dimension in non-critical string theory does not contradict with the 
$\hat c=1$ barrier \cite{DDK}. It can be understood as follows: 
fractional superstring theory has two worldsheet bosons and parafermions: 
$(\phi_L(z)$, $X_M(z)$, $\psi_L(z)$, $ \psi_M(z))$. 
Although the (matter) central charge seems to exceed the $c=1$ bound, the exceeding central charge 
(corresponding to that of parafermions $\psi_L(z)$ and $\psi_M(z)$) 
is gauged away by the fractional superconformal symmetry. 
Then we again have two-dimensional spacetime. However, as we can see in the 
type 0 superstring theory, superstring theory remains $\mathbb Z_2$ discrete spacetime 
(Stokes sectors). In the same way, as is indicated in the multi-cut matrix models, 
general $k$-fractional superstring theory would have remaining $\mathbb Z_k$ discrete 
spacetime. This extra discrete spacetime becomes the continuum third 
dimension in the $k\to \infty$ limit. In this sense, the appearance of the third dimension 
has no contradiction with the barrier of $\hat c=1$ 
in non-critical string theory (or Liouville theory).%
\footnote{One can also say that this extra dimension cannot be regard as an extra dimension 
in the Liouville theory because the neighborhood Stokes sectors ($\nu \to \nu+\delta\nu$) 
can only be observed by non-perturbative effects. }
\item In the weak coupling regime $\zeta \to \infty$, perturbative string theory description is favored and 
it is hard to see the neighborhood Stokes sectors, i.e. extra dimension. 
In the center of the spectral curve $\zeta \to 0$, 
however, the theory is strong coupling and it is easy to see the extra dimension. 
In this sense, it is natural to ask the following question: 
{\em what is the strong coupling dual theory of the $(\hat p,\hat q)$ 
minimal fractional superstring theory?} This strong coupling dual theory 
lives in the three-dimensional specetime, and controls 
all the perturbative $(\hat p,\hat q)$ minimal fractional superstring theories 
in the weak-coupling landscape. 
Therefore, it is natural to refer to this strong-coupling dual theory 
as {\em $(\hat p,\hat q)$ minimal non-critical M theory}.  
Whatever fundamental degree of freedom they have (i.e.~membrane or not), 
this three-dimensional theory is mathematically well defined and shares several features with 
the critical M theory. 
\item It is known that the multi-cut matrix models do not have good large $N$ expansion 
without fixing the filling fraction \cite{MultiCutUniversality}. 
Therefore, the strong-coupling dual theory should not have 
good expansion in terms of string coupling constant $g_{\rm str}$. 
An interesting candidate for the 
strong coupling dual description is the Kontsevich matrix models \cite{KontsevichMM} of fractional superstring theory 
because this matrix model does not rely on the string coupling topological expansion. 
\item In this multi-cut matrix-model context, the direct connection to type 0A theory is not clearly understood: 
It is because the multi-cut two-matrix models are a natural generalization of type 0B superstring theory
(i.e.~two-cut); However the usual critical M theory is directly related to type IIA superstrings. 
From the viewpoints of the orthonormal polynomial system, the difference of them is almost 
to put one eigenvalue on the origin (0A) or not (0B), and this 
introduction of eigenvalue corresponds to introduction of the D0-brane \cite{UniCom} 
(or in this context KK modes of the third dimension \cite{NonCriticalMTheory}). 
Therefore, it is interesting to consider such a 0A setting, and then 
to see to what extent the usual scenario from M theory to type IIA superstrings 
can be realized in the multi-cut matrix models. 
\item Our result indicates that each perturbative-string vacuum 
in fractional superstring theory is only a segment of the total non-perturbative theory. 
From the strong-coupling point of view, the non-perturbative vacuum of this string theory 
is a superposition of perturbative vacua. 
That is, amplitudes of full theory should include information 
of different perturbative-string vacua. 
This suggests that it is these non-perturbative vacua 
which becomes important in the string Landscape. 
Therefore, it is interesting to consider the following questions: 
{\em what is the dynamical aspects of non-perturbative vacua?} 
and {\em what is the general framework to control these vacua?}
The multi-cut matrix models should play important roles to answer these questions. 
\end{itemize}
Although the discussion in this section relies on analogies and guessworks, 
we hope that this gives a new direction of non-critical string theory, 
a solvable and non-perturbative toy model for string Landscape and possibly for M theory. 

\vspace{1cm}
\noindent
{\bf \large Acknowledgment}  
\vspace{0.2cm}

\noindent
The authors would like to thank Sheng-Yu Darren Shih 
for collaboration and various discussions in the early stage of this work. We would also like to thank
Bertrand Eynard, Kazuyuki Furuuchi, Pei-Ming Ho, Hiroshi Isono, 
Shoichi Kawamoto, Ivan Kostov, So Okada, Tomohisa Takimi, Tomohide Terasoma, Dan Tomino, 
Wen-Yu Wen and Chen-Pin Yeh for useful discussions and comments. 
HI would like to thank Pei-Ming Ho for encouragement toward the study of
``Matrix Landscape,'' which partially led him to this work, and also every people 
in IST, IPhT CEA-Saclay 
and IH\' ES for their hospitality on his visit where part of the work has been done, 
especially to Gabriel Lopes Cardoso, Maxim Kontsevich, Ivan Kostov, Didina Serban, 
Shigenori Seki and Futoshi Yagi. 
CT is supported by National Science Council of Taiwan under the contract 
No.~96-2112-M-021-002-MY3. The authors are also supported 
by National Center for Theoretical Science under NSC No.~98-2119-M-002-001.

\appendix

\section{Examples of the matrix realizations \label{ExampleMrealization}}

Here we show some examples of the matrix realizations given in \eq{MatRealComplexCases}.  
That is, several examples of $M$ ($M=M_{\pm,k}$ for the $k$-cut cases) in the realization. 
They also include other kinds of realizations which cannot be written as \eq{MatrixRealizationComplex}:

The three-cut cases:
\begin{align}
M_{+,3} = \frac{1}{\sqrt{3}}
\begin{pmatrix}
-2 & 1 & 1 \cr
1  &  1  & -2 \cr
1  & -2  & 1
\end{pmatrix}, \quad 
\frac{1}{\sqrt{3}}
\begin{pmatrix}
1 & 1 & -2 \cr
1  &  -2  & 1 \cr
-2  & 1  & 1
\end{pmatrix},\quad 
\begin{pmatrix}
-1 & 1 & 0 \cr
1  &  0 & -1 \cr
0  & -1  & 1
\end{pmatrix}. 
\end{align}
The four-cut cases have sinh and cosh solutions:
\begin{align}
M_{+,4} = 
\begin{pmatrix}
 0  & -1 &   0  & 1\cr
-1 &   0  & 1 & 0 \cr
0    & 1  &  0  & -1 \cr
1  &   0  & -1 & 0
\end{pmatrix},\qquad
M_{-,4} = 
\begin{pmatrix}
 0  & -1 &   0  & -1\cr
1 &   0  & 1 & 0 \cr
 0   & -1  &  0  & -1 \cr
1  &   0  & 1 & 0
\end{pmatrix}. \label{MatrixRealizationKeq4}
\end{align}
The five-cut case is given as 
\begin{align}
M_{+,5} = \Delta \bigl(\Gamma^0 - \Gamma^2\bigr)= 
\begin{pmatrix}
0  & -1 & 0 & 0 & 1 \cr
-1 & 0  & 0 & 1 & 0 \cr
0  &  0  & 1 & 0 & -1 \cr
0  &  1  & 0 & -1 & 0 \cr
1  &  0  & -1 & 0 & 0
\end{pmatrix}, 
\end{align}
or also 
\begin{align}
M_{+,5} = 
\begin{pmatrix}
\frac{1}{2}- \frac{3}{2\sqrt{5}} & \frac{1}{\sqrt{5}} & \frac{1}{\sqrt{5}} & \frac{1}{\sqrt{5}} & -\frac{1}{2}- \frac{3}{2\sqrt{5}} \cr
\frac{1}{\sqrt{5}} & \frac{1}{\sqrt{5}} & \frac{1}{\sqrt{5}} & -\frac{1}{2}- \frac{3}{2\sqrt{5}} & \frac{1}{2}- \frac{3}{2\sqrt{5}} \cr
\frac{1}{\sqrt{5}} & \frac{1}{\sqrt{5}} & -\frac{1}{2}- \frac{3}{2\sqrt{5}} & \frac{1}{2}- \frac{3}{2\sqrt{5}} & \frac{1}{\sqrt{5}} \cr 
\frac{1}{\sqrt{5}} & -\frac{1}{2}- \frac{3}{2\sqrt{5}} & \frac{1}{2}- \frac{3}{2\sqrt{5}} & \frac{1}{\sqrt{5}} & \frac{1}{\sqrt{5}} \cr
-\frac{1}{2}- \frac{3}{2\sqrt{5}} & \frac{1}{2}- \frac{3}{2\sqrt{5}} & \frac{1}{\sqrt{5}} & \frac{1}{\sqrt{5}} &  \frac{1}{\sqrt{5}}
\end{pmatrix}. 
\end{align}
Below the matrices $M$ in the four-cut case with real potentials are shown:
\begin{align}
M_{+,4}^{(\rm real)} = 
\begin{pmatrix}
 0  & 1 &  0  & 1\cr
1 &  0  & 1 & 0\cr
 0   & 1  &  0 & -1 \cr
1  & 0   & -1 & 0
\end{pmatrix},\qquad
M_{-,4}^{(\rm real)} = 
\begin{pmatrix}
 0  & 1 &  0  & 1\cr
-1 &  0  & -1 & 0 \cr
  0  & 1  & 0  & -1 \cr
-1  &  0  & 1 & 0
\end{pmatrix}.
\end{align}

These are the basic realizations of our solutions 
and the other realization should be related to the basic realization 
\eq{MatrixRealizationSinh} by some proper similarity transformation:
\begin{align}
M \to M' = e^{\sum_{n=0}^{k-1} t_n \Gamma^n} M e^{-\sum_{n=0}^{k-1} t_n \Gamma^n},
\end{align}
which commutes with $\Gamma$ and can give a real matrix $M'$.

\section{The $\mathbb Z_k$ charge and the free fermions  \label{ZkChargeFreeFermion}}
In this section, we consider possible implications of the relation \eq{ConnectionRelation} 
on the free-fermion formulation \cite{fi1}. 
As it has been pointed out in \cite{fi1}, {\em different eigenvalues} of the Lax operators $(\bP,\bQ)$
correspond to macroscopic loop amplitudes of FZZT-branes with {\em different charges}. 
The meaning of this statement is following:
In the free-fermion formulation, the charged FZZT branes are identified with the $k$-component free fermions 
$c_0^{(j)}(\zeta)$ $(j=1,2,\cdots,k)$. It is because the $\mathbb Z_k$ charge is measured by 
the following fermion-number operator:
\begin{align}
\text{($\mathbb Z_k$ charge)} = \sum_{j=1}^k \omega^j \, \alpha_0^{(j)},
\end{align}
where $\alpha_0^{(j)}$ is the zero-th boson oscillator of the fermion 
$c_0^{(j)}(\zeta) = :\!\!e^{\varphi_0^{(j)}(\zeta)}\!\!:$. 
In this sense, the macroscopic loop amplitudes $\mathcal Q^{(j)}(\zeta)$ 
of this charged FZZT brane should be given by the expectation value of the bosonization operator:
\begin{align}
\mathcal Q^{(j)}(\zeta) = \vev{\del_\zeta \varphi_0^{(j)}(\zeta)}. 
\end{align}
On the other hand, it was shown in \cite{fi1} that each Baker-Akhiezer function (i.e. eigenfunction) 
$\Psi^{(j)}(t;\zeta)$ of Eq.~\eq{BakAkhStokes} is written by the free-fermion operator 
$c_0^{(j)}(\zeta)$ of multi-component KP hierarchy in the following way:%
\footnote{See Eq.(3.11) of \cite{fi1}, and its derivations. }
\begin{align}
\Psi^{(j)}(t;\zeta) = 
\begin{pmatrix}
\Psi^{(j)}_1(t;\zeta) \cr 
\Psi^{(j)}_2(t;\zeta) \cr 
\vdots \cr
\Psi^{(j)}_k(t;\zeta)
\end{pmatrix}, \qquad
\Psi^{(j)}_l (t;\zeta)= \frac{\bra{t/g_{\rm str}}e^{-\phi^{(l)}}c_0^{(j)}( \omega^{-(j-1)}\zeta) \ket{\Phi}}
{\bracket{t/g_{\rm str}}{\Phi}}. \label{BakAkhFermion}
\end{align}
Since difference of index $l$ in $\Psi^{(j)}_l(t;\zeta)$ is difference of background flux 
and negligible in the week coupling limit \cite{SeSh2,fi1}, the behavior is uniformly given as
\begin{align}
\Psi^{(j)}(t; \omega^{j-1}\zeta) \sim \exp\Bigl[ \frac{1}{g_{\rm str}}\, \int^{\zeta} \, d\zeta' \, 
\mathcal Q^{(j)}(\zeta') \Bigr]\,
\chi^{(j)}(t;\zeta),
\end{align}
in the week coupling limit $g_{\rm str} \to 0$. 
Therefore, the macroscopic loop amplitudes, $\mathcal Q^{(j)}(\zeta)$, of FZZT brane of a charge $j$ 
are identified with the eigenvalues of the Lax operators as 
\begin{align}
\mathcal Q^{(j)}(\zeta) \equiv  \vev{\del_\zeta \varphi_0^{(j)}(\zeta)} 
= \omega^{j-1}Q^{(j)}(\omega^{j-1}\zeta). 
\end{align}
However, since we now know that the Baker-Akhiezer functions are connected to each other, 
one should say that each fermion operators should be restricted in each Stokes sector:
\begin{align}
c_0^{(j)}(\zeta)  \qquad \text{in \,
$2\pi \nu_j -\frac{\pi }{k} < \arg (\zeta) <  2\pi \nu_j +\frac{\pi }{k}$}, 
\end{align}
otherwise the fermion operator should be renamed:
\begin{align}
c_0^{(j)}(\zeta) \equiv  c_0^{(j')}(\zeta) \qquad 
\text{in \,
$2\pi \nu_j -\frac{\pi }{k} < \arg (\zeta) <  2\pi \nu_j +\frac{\pi }{k}$}, \label{ChangeOfName}
\end{align}
in the non-perturbative sense. 

This means that there is only one (non-perturbative) fundamental FZZT brane (free fermion), 
$c_0(\zeta;\nu)$, in this three-dimensional spacetime and 
that the the charge $\nu$ of FZZT branes is just a coordinate which 
measures which Stokes sector the FZZT brane exists: $c_0(\zeta;\nu)= c_0^{(\nu)}( e^{-2\pi i \nu} \zeta)$. 
This is quite natural because originally there is only one kind of eigenvalues 
in the matrix models and the position of eigenvalues (which Fermi sea the eigenvalue exists) 
is interpreted as the charge of D-brane \cite{TT,NewHat}. 

Finally, we note an interesting expression of the Baker-Akhiezer function $\Psi^{(j)}(t;\zeta)$. 
According to Eq.~\eq{BakAkhFermion}, 
$\Psi^{(j)}_j(t;\zeta)$ component gives simple one-point function of the FZZT brane with 
charge $j$: $\Psi^{(j)}_j(t;\zeta) = \vev{c_0^{(j)}(\omega^{-(j-1)}\zeta)}$. This means that one has 
the following expression:
\begin{align}
\Psi_l(t;\zeta) = \vev{c_0^{(l)}(\omega^{-(l-1)}\zeta)}. 
\end{align}
in the sense of Eq.~\eq{ChangeOfName}. 
Therefore, amplitudes of FZZT brane with a charge $j$ are obtained by 
properly analytic continuing the $j$-th component of the orthonormal polynomials $\Psi(t;\zeta)$. 
This is the observation given in \cite{SeSh2}.


\begin{thebibliography}{99}

\bibitem{Polyakov}
  A.~M.~Polyakov,
   ``Quantum geometry of bosonic strings,''
  Phys.\ Lett.\ B {\bf 103} (1981) 207; 
   ``Quantum geometry of fermionic strings,''
  Phys.\ Lett.\ B {\bf 103} (1981) 211.

\bibitem{KPZ}
  V.~G.~Knizhnik, A.~M.~Polyakov and A.~B.~Zamolodchikov,
   ``Fractal structure of 2d-quantum gravity,''
  Mod.\ Phys.\ Lett.\ A {\bf 3} (1988) 819.

\bibitem{DDK}
  F.~David,
   ``Conformal field theories coupled to 2-D gravity in the conformal gauge,''
  Mod.\ Phys.\ Lett.\ A {\bf 3} (1988) 1651;\\
  J.~Distler and H.~Kawai,
   ``Conformal field theory and 2-D quantum gravity, 
    or who's afraid of Joseph Liouville?,''
  Nucl.\ Phys.\ B {\bf 321} (1989) 509.

\bibitem{DOZZ}
H.~Dorn and H.~J.~Otto,
``Two and three point functions in Liouville theory,''
Nucl.\ Phys.\ B {\bf 429} (1994) 375
[arXiv:hep-th/9403141]; \\
A.~B.~Zamolodchikov and Al.~B.~Zamolodchikov,
``Structure constants and conformal bootstrap in Liouville field theory,''
Nucl.\ Phys.\ B {\bf 477} (1996) 577
[arXiv:hep-th/9506136]. 

\bibitem{Teschner}
  J.~Teschner,
  ``On the Liouville three point function,''
  Phys.\ Lett.\  B {\bf 363} (1995) 65
  [arXiv:hep-th/9507109].


\bibitem{FZZT}
V.~Fateev, A.~B.~Zamolodchikov and Al.~B.~Zamolodchikov,
``Boundary Liouville field theory. I: 
Boundary state and boundary two-point function,''
arXiv:hep-th/0001012;\\
%
J.~Teschner,
``Remarks on Liouville theory with boundary,''
arXiv:hep-th/0009138.

\bibitem{ZZ}
  A.~B.~Zamolodchikov and Al.~B.~Zamolodchikov,
  ``Liouville field theory on a pseudosphere,''
  arXiv:hep-th/0101152.

\bibitem{sDOZZ}
  R.~H.~Poghosian,
  ``Structure constants in the $N = 1$ super-Liouville field theory,''
  Nucl.\ Phys.\  B {\bf 496} (1997) 451
  [arXiv:hep-th/9607120];\\
%
  R.~C.~Rashkov and M.~Stanishkov,
  ``Three-point correlation functions in $N=1$ Super Liouville Theory,''
  Phys.\ Lett.\  B {\bf 380} (1996) 49
  [arXiv:hep-th/9602148]. 
  
\bibitem{fuku-hoso}
  T.~Fukuda and K.~Hosomichi,
  ``Super Liouville theory with boundary,''
  Nucl.\ Phys.\ B {\bf 635} (2002) 215
  [arXiv:hep-th/0202032]; \\
  C.~Ahn, C.~Rim and M.~Stanishkov,
  ``Exact one-point function of $N = 1$ super-Liouville theory with boundary,''
  Nucl.\ Phys.\  B {\bf 636} (2002) 497
  [arXiv:hep-th/0202043].

\bibitem{DSL}
  E.~Brezin and V.~A.~Kazakov,
   ``Exactly solvable field theories of closed strings,''
  Phys.\ Lett.\  B {\bf 236} (1990) 144;\\
  M.~R.~Douglas and S.~H.~Shenker,
  Nucl.\ Phys.\  B {\bf 335} (1990) 635;\\
  D.~J.~Gross and A.~A.~Migdal,
   ``Nonperturbative Two-Dimensional Quantum Gravity,''
  Phys.\ Rev.\ Lett.\  {\bf 64} (1990) 127.

\bibitem{TwoMatString}
  E.~Brezin, M.~R.~Douglas, V.~Kazakov and S.~H.~Shenker,
   ``The Ising model coupled to 2-d Gravity: A nonperturbative analysis,''
  Phys.\ Lett.\  B {\bf 237} (1990) 43;\\
  D.~J.~Gross and A.~A.~Migdal,
   ``Nonperturbative Solution of the Ising Model on a Random Surface,''
  Phys.\ Rev.\ Lett.\  {\bf 64} (1990) 717.

\bibitem{GrossMigdal2}
  D.~J.~Gross and A.~A.~Migdal,
   ``A nonperturbative treatment of two-dimensional quantum gravity,''
  Nucl.\ Phys.\  B {\bf 340} (1990) 333.

\bibitem{fkn}
  M.~Fukuma, H.~Kawai and R.~Nakayama,
  ``Continuum Schwinger-Dyson equations and universal structures 
   in two-dimensional quantum gravity,''
  Int.\ J.\ Mod.\ Phys.\ A {\bf 6} (1991) 1385;
  ``Infinite dimensional Grassmannian structure 
   of two-dimensional quantum gravity,''
  Commun.\ Math.\ Phys.\  {\bf 143} (1992) 371;
  ``Explicit solution for $p$--$q$ duality
   in two-dimensional quantum gravity,''
  Commun.\ Math.\ Phys.\  {\bf 148} (1992) 101.

\bibitem{fy12}
  M.~Fukuma and S.~Yahikozawa,
  ``Nonperturbative effects in noncritical strings with soliton 
  backgrounds,''
  Phys.\ Lett.\ B {\bf 396} (1997) 97
  [arXiv:hep-th/9609210];
  ``Combinatorics of solitons in noncritical string theory,''
  Phys.\ Lett.\ B {\bf 393} (1997) 316
  [arXiv:hep-th/9610199];
  %

\bibitem{fy3}
  M.~Fukuma and S.~Yahikozawa,
  ``Comments on D-instantons in c $<$ 1 strings,''
  Phys.\ Lett.\  B {\bf 460} (1999) 71
  [arXiv:hep-th/9902169].

\bibitem{Shenker}
  S.~H.~Shenker,
  ``The Strength of nonperturbative effects in string theory,''

\bibitem{Polchinski}
  J.~Polchinski,
  ``Combinatorics of boundaries in string theory,''
  Phys.\ Rev.\  D {\bf 50} (1994) 6041
  [arXiv:hep-th/9407031].
  
\bibitem{Reloaded}
  J.~McGreevy and H.~L.~Verlinde,
  ``Strings from tachyons: The c = 1 matrix reloaded,''
  JHEP {\bf 0312} (2003) 054
  [arXiv:hep-th/0304224].

\bibitem{Paradigm}
  D.~Gaiotto and L.~Rastelli,
  ``A paradigm of open/closed duality: Liouville D-branes and the  Kontsevich
  model,''
  JHEP {\bf 0507} (2005) 053
  [arXiv:hep-th/0312196].

 \bibitem{GrossWitten}
   D.~J.~Gross and E.~Witten,
    ``Possible Third Order Phase Transition In The Large N Lattice Gauge
   Theory,''
   Phys.\ Rev.\  D {\bf 21} (1980) 446.

 \bibitem{PeShe}
   V.~Periwal and D.~Shevitz,
    ``Unitary matrix models as exactly solvable string theories,''
   Phys.\ Rev.\ Lett.\  {\bf 64} (1990) 1326;
    ``Exactly solvable unitary matrix models: multicritical potentials 
       and correlations,''
   Nucl.\ Phys.\ B {\bf 344} (1990) 731.

 \bibitem{DSS}
   M.~R.~Douglas, N.~Seiberg and S.~H.~Shenker,
    ``Flow and instability in quantum gravity,''
   Phys.\ Lett.\  B {\bf 244} (1990) 381.

 \bibitem{Nappi}
   C.~R.~Nappi,
    ``Painleve-II And Odd Polynomials,''
   Mod.\ Phys.\ Lett.\  A {\bf 5} (1990) 2773.

 \bibitem{CDM}
   C.~Crnkovic, M.~R.~Douglas and G.~W.~Moore,
    ``Loop equations and the topological phase of multi-cut matrix models,''
   Int.\ J.\ Mod.\ Phys.\ A {\bf 7} (1992) 7693
   [arXiv:hep-th/9108014].

\bibitem{HMPN}
  T.~J.~Hollowood, L.~Miramontes, A.~Pasquinucci and C.~Nappi,
   ``Hermitian Versus Anti-Hermitian One Matrix Models And Their Hierarchies,''
  Nucl.\ Phys.\  B {\bf 373} (1992) 247
  [arXiv:hep-th/9109046].

\bibitem{TT}
  T.~Takayanagi and N.~Toumbas,
  ``A matrix model dual of type 0B string theory in two dimensions,''
  JHEP {\bf 0307} (2003) 064
  [arXiv:hep-th/0307083].

\bibitem{NewHat}
  M.~R.~Douglas, I.~R.~Klebanov, D.~Kutasov, J.~M.~Maldacena, E.~J.~Martinec and N.~Seiberg,
  ``A new hat for the $c = 1$ matrix model,''
  arXiv:hep-th/0307195.
  
\bibitem{UniCom}
  I.~R.~Klebanov, J.~M.~Maldacena and N.~Seiberg,
   ``Unitary and complex matrix models as 1-d type 0 strings,''
  Commun.\ Math.\ Phys.\  {\bf 252} (2004) 275
  [arXiv:hep-th/0309168].
  
\bibitem{irie2}
  H.~Irie,
   ``Fractional supersymmetric Liouville theory and the multi-cut matrix models,''
  Nucl.\ Phys.\ B {\bf 819} (2009) 351 
  [arXiv:0902.1676 [hep-th]]. 
  
\bibitem{MultiCut}
  C.~Crnkovic and G.~W.~Moore,
   ``Multicritical multicut matrix models,''
  Phys.\ Lett.\  B {\bf 257} (1991) 322.
 
\bibitem{FSST}
  P.~C.~Argyres, A.~LeClair and S.~H.~H.~Tye,
  ``On The Possibility Of Fractional Superstrings,''
  Phys.\ Lett.\  B {\bf 253} (1991) 306.

\bibitem{ZF}
  A.~B.~Zamolodchikov and V.~A.~Fateev,
  ``Parafermionic Currents In The Two-Dimensional Conformal Quantum Field
  Theory And Selfdual Critical Points In Z(N) Invariant Statistical Systems,''
  Sov.\ Phys.\ JETP {\bf 62} (1985) 215
  [Zh.\ Eksp.\ Teor.\ Fiz.\  {\bf 89} (1985) 380].


\bibitem{fi1}
  M.~Fukuma and H.~Irie,
   ``A string field theoretical description of $(p,q)$ minimal superstrings,''
  JHEP {\bf 0701} (2007) 037
  [arXiv:hep-th/0611045].
  
\bibitem{kcKP}
M.~Sato, 
RIMS Kokyuroku {\bf 439} (1981) 30; \\
%
  E.~Date, M.~Jimbo, M.~Kashiwara and T.~Miwa,
   ``Transformation groups for soliton equations. 3. 
     Operator approach to the Kadomtsev-Petviashvili equation,''
RIMS-358; \\
  M.~Jimbo and T.~Miwa,
  ``Solitons and infinite dimensional Lie algebras,''
  Publ.\ Res.\ Inst.\ Math.\ Sci.\ Kyoto {\bf 19} (1983) 943; \\
  V.~G.~Kac and J.~W.~van de Leur,
  ``The $n$-component KP hierarchy and representation theory,''
  J.\ Math.\ Phys.\  {\bf 44} (2003) 3245
  [arXiv:hep-th/9308137].

\bibitem{CISY1}
  C.~T.~Chan, H.~Irie, S.~Y.~Shih and C.~H.~Yeh,
  ``Macroscopic loop amplitudes in the multi-cut two-matrix models,''
  Nucl.\ Phys.\  B {\bf 828} (2010) 536
  [arXiv:0909.1197 [hep-th]].

\bibitem{BIPZ}
  E.~Brezin, C.~Itzykson, G.~Parisi and J.~B.~Zuber,
  ``Planar Diagrams,''
  Commun.\ Math.\ Phys.\  {\bf 59} (1978) 35.

\bibitem{KazakovSeries}
  V.~A.~Kazakov,
  ``The Appearance of Matter Fields from Quantum Fluctuations of 2D Gravity,''
  Mod.\ Phys.\ Lett.\  A {\bf 4} (1989) 2125.

\bibitem{Kostov1}
  I.~K.~Kostov,
  ``Strings embedded in Dynkin diagrams,''
  Cargese 1990, Proceedings, Random surfaces and quantum gravity, pp.135-149. 

\bibitem{Kostov2}
  I.~K.~Kostov,
  ``Loop amplitudes for nonrational string theories,''
  Phys.\ Lett.\  B {\bf 266} (1991) 317.

\bibitem{Kostov3}
  I.~K.~Kostov,
  ``Strings with discrete target space,''
  Nucl.\ Phys.\  B {\bf 376} (1992) 539
  [arXiv:hep-th/9112059].

\bibitem{BDSS}
  T.~Banks, M.~R.~Douglas, N.~Seiberg and S.~H.~Shenker,
  ``Microscopic and macroscopic loops in nonperturbative two-dimensional gravity,''
  Phys.\ Lett.\  B {\bf 238} (1990) 279.

\bibitem{MSS}
  G.~W.~Moore, N.~Seiberg and M.~Staudacher,
  ``From loops to states in 2-D quantum gravity,''
  Nucl.\ Phys.\  B {\bf 362} (1991) 665.

\bibitem{DKK}
  J.~M.~Daul, V.~A.~Kazakov and I.~K.~Kostov,
   ``Rational theories of 2-D gravity from the two matrix model,''
  Nucl.\ Phys.\  B {\bf 409} (1993) 311
  [arXiv:hep-th/9303093].

\bibitem{Kris}
  C.~Kristjansen,
  ``Multiloop correlators for rational theories of 2-d gravity from the
  generalized Kontsevich models,''
  Nucl.\ Phys.\  B {\bf 436} (1995) 342
  [arXiv:hep-th/9409066].

\bibitem{AnazawaIshikawaItoyama1}
  M.~Anazawa, A.~Ishikawa and H.~Itoyama,
  ``Universal annulus amplitude from the two matrix model,''
  Phys.\ Rev.\  D {\bf 52} (1995) 6016
  [arXiv:hep-th/9410015].

\bibitem{AnazawaIshikawaItoyama2}
  M.~Anazawa, A.~Ishikawa and H.~Itoyama,
  ``Macroscopic three loop amplitudes from the two matrix model,''
  Phys.\ Lett.\  B {\bf 362} (1995) 59
  [arXiv:hep-th/9508009].

\bibitem{AnazawaItoyama}
  M.~Anazawa and H.~Itoyama,
  ``Macroscopic $n$-Loop Amplitude for Minimal Models Coupled to
  Two-Dimensional Gravity: Fusion Rules and Interactions,''
  Nucl.\ Phys.\  B {\bf 471} (1996) 334
  [arXiv:hep-th/9511220].


\bibitem{David}  F.~David,
  ``Phases of the large N matrix model and nonperturbative effects in 2-d gravity,''
  Nucl.\ Phys.\  B {\bf 348} (1991) 507;
  ``Nonperturbative effects in matrix models and vacua of two-dimensional
  Phys.\ Lett.\  B {\bf 302} (1993) 403
  [arXiv:hep-th/9212106].

\bibitem{KazakovKostov}
  V.~A.~Kazakov and I.~K.~Kostov,
  ``Instantons in non-critical strings from the two-matrix model,''
  arXiv:hep-th/0403152.

\bibitem{SeSh}
  N.~Seiberg and D.~Shih,
   ``Branes, rings and matrix models in minimal (super)string theory,''
  JHEP {\bf 0402} (2004) 021
  [arXiv:hep-th/0312170].
  
\bibitem{MMSS}
  J.~M.~Maldacena, G.~W.~Moore, N.~Seiberg and D.~Shih,
   ``Exact vs. semiclassical target space of the minimal string,''
  JHEP {\bf 0410} (2004) 020
  [arXiv:hep-th/0408039].

\bibitem{SeSh2}
  N.~Seiberg and D.~Shih,
   ``Flux vacua and branes of the minimal superstring,''
  JHEP {\bf 0501} (2005) 055
  [arXiv:hep-th/0412315].

\bibitem{KOPSS}
  D.~Kutasov, K.~Okuyama, J.~w.~Park, N.~Seiberg and D.~Shih,
  ``Annulus amplitudes and ZZ branes in minimal string theory,''
  JHEP {\bf 0408} (2004) 026
  [arXiv:hep-th/0406030].
  
\bibitem{fim}
  M.~Fukuma, H.~Irie and Y.~Matsuo,
   ``Notes on the algebraic curves in $(p,q)$ minimal string theory,''
  JHEP {\bf 0609} (2006) 075
  [arXiv:hep-th/0602274].
  
\bibitem{NonCriticalMTheory}
  P.~Horava and C.~A.~Keeler,
  ``Noncritical M-theory in 2+1 dimensions as a nonrelativistic Fermi liquid,''
  JHEP {\bf 0707} (2007) 059
  [arXiv:hep-th/0508024].

\bibitem{Mehta}
  M.~L.~Mehta,
  ``A Method Of Integration Over Matrix Variables,''
  Commun.\ Math.\ Phys.\  {\bf 79}, 327 (1981).

\bibitem{TadaYamaguchiDouglas}
  T.~Tada and M.~Yamaguchi,
   ``$P$ and $Q$ operator analysis for two matrix model,''
  Phys.\ Lett.\  B {\bf 250} (1990) 38; \\
  M.~R.~Douglas,
   ``The Two matrix model,''
{\it  In *Cargese 1990, Proceedings, Random surfaces and quantum gravity* 77-83. (see HIGH ENERGY PHYSICS INDEX 30 (1992) No. 17911)}; \\
  T.~Tada,
   ``$(Q,P)$ Critical Point From Two Matrix Models,''
  Phys.\ Lett.\  B {\bf 259} (1991) 442.

\bibitem{DouglasGeneralizedKdV}
  M.~R.~Douglas,
   ``Strings in less than one-dimension and the generalized KdV hierarchies,''
  Phys.\ Lett.\  B {\bf 238} (1990) 176.
  
\bibitem{Moore}
  G.~W.~Moore,
  ``Geometry Of The String Equations,''
  Commun.\ Math.\ Phys.\  {\bf 133} (1990) 261;
  ``Matrix Models Of 2-D Gravity And Isomonodromic Deformation,''
  Prog.\ Theor.\ Phys.\ Suppl.\  {\bf 102} (1990) 255.

\bibitem{EynardZinnJustin}
  B.~Eynard and J.~Zinn-Justin,
  ``Large order behavior of 2-D gravity coupled to $d < 1$ matter,''
  Phys.\ Lett.\  B {\bf 302} (1993) 396
  [arXiv:hep-th/9301004].
  
\bibitem{Irie1}
  H.~Irie,
  ``Notes on D-branes and dualities in $(p,q)$ minimal superstring theory,''
Nucl.\ Phys.\  B {\bf 794} [PM] (2008) 402
  [arXiv:0706.4471 hep-th].

\bibitem{NashDet}
  P.~L.~Nash,
  ``Chebyshev Polynomials and Quadratic Path Integrals,'' 
  J. Math. Phys. 27 (1986) 2963.

\bibitem{EynardAlg}
  B.~Eynard,
  ``Large N expansion of the 2-matrix model,''
  JHEP {\bf 0301} (2003) 051
  [arXiv:hep-th/0210047].

\bibitem{BakAkhNonPerTwoCut}
  A.~R.~Its and V.~Y.~Novokshenov, ``The Isomonodromic Deformation Method in the
Theory of Painlev\' e Equations,'' Springer-Verlag (1986);\\
  P.~Bleher and A.~Its,
  ``Double scaling limit in the random matrix model: the Riemann-Hilbert
  approach,''
  arXiv:math-ph/0201003.

\bibitem{fis}
  M.~Fukuma, H.~Irie and S.~Seki,
  ``Comments on the D-instanton calculus in $(p,p+1)$ minimal string theory,''
  Nucl.\ Phys.\  B {\bf 728} (2005) 67
  [arXiv:hep-th/0505253].

\bibitem{KaluzaKlein}
  T.~Kaluza,
  ``On The Problem Of Unity In Physics,''
  Sitzungsber.\ Preuss.\ Akad.\ Wiss.\ Berlin (Math.\ Phys.\ ) {\bf 1921} (1921) 966;\\
  O.~Klein,
  ``Quantum theory and five-dimensional theory of relativity,''
  Z.\ Phys.\  {\bf 37} (1926) 895
  [Surveys High Energ.\ Phys.\  {\bf 5} (1986) 241].



\bibitem{MultiCutUniversality}
  G.~Bonnet, F.~David and B.~Eynard,
  ``Breakdown of universality in multi-cut matrix models,''
  J.\ Phys.\ A  {\bf 33} (2000) 6739
  [arXiv:cond-mat/0003324].

\bibitem{GinspargZinnJustin}
  P.~H.~Ginsparg and J.~Zinn-Justin,
  ``Action principle and large order behavior of nonperturbative gravity,''

\bibitem{DiFKutasov}
   P.~Di Francesco and D.~Kutasov,
  ``World Sheet and Space Time Physics in Two Dimensional (Super) String
  Theory,''
  Nucl.\ Phys.\  B {\bf 375} (1992) 119
  [arXiv:hep-th/9109005].

\bibitem{Klebanov}
  I.~R.~Klebanov,
  ``String Theory In Two-Dimensions,''
  arXiv:hep-th/9108019.

\bibitem{EynardLoop}
  B.~Eynard,
  ``Topological expansion for the 1-hermitian matrix model correlation
  functions,''
  JHEP {\bf 0411} (2004) 031
  [arXiv:hep-th/0407261];\\
  B.~Eynard and N.~Orantin,
  ``Invariants of algebraic curves and topological expansion,''
  arXiv:math-ph/0702045;
  ``Geometrical interpretation of the topological recursion, and integrable
  string theories,''
  arXiv:0911.5096 [math-ph].

\bibitem{KontsevichMM}
  M.~Kontsevich,
  ``Intersection theory on the moduli space of curves and the matrix Airy
  function,''
  Commun.\ Math.\ Phys.\  {\bf 147} (1992) 1.




\end{thebibliography}
\end{document}